\documentclass[pra,twocolumn,showpacs,preprintnumbers,superscriptaddress,
amsmath,amssymb,tightenlines,epsfig]{revtex4}

\usepackage{bm}% bold math
\usepackage{graphicx}% Include figure files

\newcommand{\rv}{{\bf r}}
\newcommand{\Ev}{{\bf E}}

\newcommand{\dv}{{\bf d}}

\newcommand{\beq}{\begin{equation}}
\newcommand{\eeq}{\end{equation}}
\newcommand{\bea}{\begin{eqnarray}}
\newcommand{\eea}{\end{eqnarray}}

\renewcommand{\>}{\rangle}
\renewcommand{\(}{\left(}
\renewcommand{\)}{\right)}
\renewcommand{\[}{\left[}
\renewcommand{\]}{\right]}
\newcommand{\commentout}[1]{{}}

\begin{document}

\title{Engineering vortex rings and systems for controlled studies of vortex\\ interactions in
Bose-Einstein condensates}
\author{Janne Ruostekoski}
\affiliation{Department of Physics, Astronomy and Mathematics,
University of Hertfordshire, Hatfield, Herts, AL10 9AB, UK}
%\affiliation{Institute for Theoretical Atomic and Molecular Physics,
%Harvard-Smithsonian Center for Astrophysics, Cambridge MA 02135}
%\email{j.ruostekoski@herts.ac.uk}
\author{Zachary Dutton}
\affiliation{Naval Research Laboratory, Washington, DC 20375}

\begin{abstract}
We study controlled methods of preparing vortex configurations in
atomic Bose-Einstein condensates and their use in the studies of
fundamental vortex scattering, reconnection processes and superfluid
sound emission. We explore techniques of imprinting closed vortex
rings by means of coherently driving internal atomic transitions
with electromagnetic fields which exhibit singular phase profiles.
In particular, we show that a vortex ring can be prepared in a
controlled way by two focused co-propagating Gaussian laser beams.
More complex vortex systems may also be imprinted by directly
superposing simpler field configurations or by programming their
phase profiles on optical holograms. This provides the controlled
method of studying vortex reconnections in atomic superfluids.  We
analyze specific examples of two merging vortex rings in a trapped
two-component $^{87}$Rb condensate. We calculate the radiated sound
energy in the vortex ring reconnection process and show that the
vortex relaxation and the re-distribution of sound energy can be
controlled by the imprinting process. The energy is first
concentrated towards the trap center and later emitted outwards as
sound and transformed to surface excitations. As another such
technique, we study creating pairs of 2D point vortices in
Bose-Einstein condensates using a `light roadblock' in ultra-slow
light propagation. We show how this can be used to study vortex
collisions in compressible superfluids and how these collisions
result in energy dissipation via phonons and, sometimes,
annihilation of vortex pairs.
\end{abstract}

\date{\today}
\maketitle

\section{Introduction}

One of the advantages of atomic Bose-Einstein condensates (BECs), as
compared to more traditional quantum fluids, is the dramatic
flexibility of experimental preparation. Both the internal and the
center-of-mass states of ultra-cold atoms can be manipulated and
controlled in diverse ways using electromagnetic (em) fields. In
this paper we explore the possibilities of using the present day
atomic physics technology as a sophisticated state engineering tool,
in order to construct highly non-trivial superfluid states in atomic
many-particle systems. In particular, such a `field-theory
engineering' may be useful in preparing topological defects,
studying their interactions, and the decay of superfluid turbulence
\cite{DON91}. We consider em field configurations which exhibit
phase singularities in the field amplitude. When such a field is
used to drive internal atomic transitions, the non-uniform phase
profile of the em field amplitude may be imprinted on the matter
wave, resulting in a topologically non-trivial phase profile in the
atomic condensate. We show how a reasonably simple em field
configuration of two phase-coherent Gaussian laser beams could be
used to imprint a vortex ring in atomic BECs by simultaneously
controlling the position, the orientation, and the radius of the
ring. More complex vortex structures, such as knotted vortex lines,
may be imprinted using optical holograms.

Moreover, we study in detail the application of this technique to
the imprinting of a pair of vortex rings and study their
reconnection dynamics in a trapped two-component $^{87}$Rb
condensate, using experimentally feasible parameters. Such
reconnections have escaped experimental observation in BECs to date.
The two rings merge together and the resulting turbulent dynamics
generates an effective local dissipation in the gas, allowing the
vortex configuration to relax to lower energy without an explicit
damping term in the dynamics. This mechanism is also analogous to
the dynamical formation of a vortex lattice in a rotating
zero-temperature BEC \cite{LOB04}. We qualitatively analyze the
re-distribution of sound energy in the trapped cloud due to the
vortex reconnection processes. Here the excitation of Kelvin waves
and the radiated sound energy can be controlled by changing the
parameters of the em fields. Although vortex rings, as localized
singular defects which are detached from superfluid boundaries, are
interesting in their own right, they can also form a building block
of energetically stable localized particle-like solitons (3D
Skyrmions) in a two-component $^{87}$Rb BEC
\cite{RUO01,SAV03b,BAT02}, providing a link between ultra-cold atom
physics, elementary particles, and cosmology.

As another method of engineering systems for studying fundamental
vortex interactions, we consider the generation of a `light
roadblock' \cite{DUT01} by abruptly distorting the ultra-slow light
propagation \cite{Nature1,KAS99,BUD99} in a BEC. The strong
nonlinear coherent light-matter coupling is here based on the method
of electromagnetically-induced transparency (EIT) \cite{EIT}, which
allows propagation of light pulses through the BEC due to the lack
of absorption. If the probe and the coupling beams are propagating
in orthogonal directions and the coupling field is varied quickly to
zero over a short distance near the BEC center, the probe field is
abruptly compressed and stopped inside the atom cloud.  This results
in large transfer of atom population between the internal states and
the emergence of a very narrow density defect inside the BEC, which
subsequently generates solitons via quantum shock waves \cite{DUT01}
and then vortex pairs via the snake instability
\cite{KP70,FED00,AND01}.  In Ref.~\cite{DUTEPN} it was noted that
this can be used to create a `gas' of multiple vortex particles, of
both circulations, which are out of equilibrium and subsequently
interact. Here we show that in a 2D geometry the generation of the
defects can be controlled to yield desired configurations of point
vortex pairs whose collisions and sound emission can be investigated
in detail. In particular, we focus on a regime in which the size of
the vortex pairs is comparable to the healing length in the
superfluid. In such a case, collisions of vortex pairs are more
complicated than they are in the imcompressible limit \cite{FET99},
as the collisions can cause energy to be dissipated in the form of
phonons.  This dissipation alters the collision dynamics and can
lead to annihilations of vortex pairs, another interesting
phenomenon which has hitherto been unobserved experimentally.  We
present numerical study of this and show how the light roadblock may
allow its observation.

The quantized vorticity in atomic superfluids has inspired
considerable experimental and theoretical activity in recent years
\cite{ANG}. There have been several theoretical proposals to imprint
vortex line singularities on atomic BECs by means of transferring
angular momentum on atoms from em fields
\cite{BOL98,MAR97,DUM98,DOB99,WIL99,RUO00b,ISO00,PU01,DUT04} and
some of the techniques have already been experimentally realized
\cite{MAT99,LEA02,LEA03,SHI04}. In Ref.~\cite{MAT99}, a vortex line
with one unit of circulation was imprinted on a pair of condensates
occupying different internal levels of $^{87}$Rb using a Raman
coupling which was set locally resonant in a small region by rapidly
rotating laser beams. In Refs.~\cite{LEA02,LEA03,SHI04}, singular
singly-quantized and doubly-quantized, as well as non-singular
coreless vortex lines were created by adiabatically inverting the
magnetic bias field along the trap axis. Also dark solitons have
been imprinted on atomic BECs, e.g., by imaging the atom cloud
through an absorption plate \cite{BUR99,DEN00}. Such $\pi$ phase
kink planes have been observed to decay into a hybrid of vortex
rings and lines through the dynamical `snake' instability
\cite{AND01,DUT01,GIN05}. A controlled method of creating vortex
rings and particlelike solitons was proposed in Ref.~\cite{RUO01},
where it was shown that a topological phase singularity forming a
closed circular loop may be imprinted on the matter field while
changing the internal state of the atoms. Ring defects in spinor
BECs may also form from simpler core structures as a result of
dissipation \cite{RUO03}. The defect generation using a `light
roadblock' was first realized in Ref.~\cite{DUT01}. More complicated
defect configurations were recently produced by superposing multiple
roadblocks \cite{GIN05}. It has also been proposed that the strong
light-matter coupling of the ultra-slow and stopped light
\cite{Nature2,OtherStoppedLight,stoppedLightTheory} could be used to
efficiently transfer vortex states between light and atomic BECs and
to store light modes with orbital angular momentum inside the BECs
\cite{DUT04}, as well as to use a vortex lattice to create a
photonic band gap \cite{OKT05}.

In Section~\ref{rings} we show how em fields containing phase
singularities can be used to imprint vortex rings on BECs.  We
discuss several examples of how the required fields can be generated
by superpositions of plane and/or Gaussian laser fields. In
Section~\ref{reconnections} we apply the method to generating two
vortex rings and studying their ensuing reconnection dynamics.  In
Section~\ref{slow} we then turn using the light roadblock to
generate of systems of multiple vortex point particles in a 2D
geometry, focusing on collisions of vortex pairs, which dissipate
energy via phonons and sometimes leads to vortex annihilations.

\section{Controlled preparation of vortex rings}
\label{rings}

\subsection{Coupled two-component condensate}

A vortex ring can be engineered by using an em field to imprint
topological phase singularities on the matter field while changing
the internal state of the atoms \cite{RUO01}. Here we consider a
two-component BEC where the two internal states are coupled by means
of the em fields with the Rabi frequency, $\Omega({\bf r})$. The
dynamics of the BECs with the Rabi coupling between the levels
$|i\>$ and $|j\>$ follows from the coupled Gross-Pitaevskii equation
(GPE)
\begin{equation}
i \hbar {\partial \psi_i\over \partial t} = ( H_0+\delta_i +\sum_k
\kappa_{ik} |\psi_k|^2 ) \psi_i+\hbar\Omega^* \psi_j\,. \label{gpe}
\end{equation}
We assume the two-component condensate, with the total number of $N$
atoms and the atomic mass $m$, to be confined in a perfectly
overlapping, isotropic trap with the trap frequency $\omega$:
\begin{equation}
H_0\equiv -{\frac{\hbar^2}{2m}}{\bf \nabla}^2+{\frac{1}{2}}
m\omega^2 r^2\,.
\end{equation}
The parameter $\delta_j(\rv)$ incorporates the detuning of the em
fields from the resonance of the internal transition as well as em
field-induced level shifts which may be non-uniform in the case of a
spatially varying field intensity. We have also defined the
interaction coefficients $\kappa_{ij}\equiv 4\pi\hbar^2 a_{ij}N/m$,
with the intraspecies and the inter species scattering lengths
denoted by $a_{ii}$ and $a_{ij}$ ($i\neq j$), respectively. Such a
system has been experimentally realized using, e.g., the $|2\>\equiv
|S_{1/2},F=2,M_F=+1\>$ and $|1\>\equiv |S_{1/2},F=1,M_F=-1\>$
hyperfine spin states of $^{87}$Rb. The two hyperfine levels were
coupled by means of a two-photon transition (one microwave and one
rf photon). For the $^{87}$Rb components the interaction strengths
are nearly equal, with $a_{11}:a_{21}:a_{22}::1.03:1:0.97$ and
$a_{21}\simeq5.50$~nm \cite{HALL98}. Since the scattering lengths
satisfy $a_{21}^2\agt a_{11}a_{22}$, the two species experience
dynamical phase separation and can strongly repel each other
\cite{HALL98}. In the absence of the em coupling between the
different components the interatomic interactions of the two
$^{87}$Rb components do not mix the atom population and the atom
numbers of the both species are separately conserved.

The phase profile of the driving em field can be imprinted on the
matter wave by means of transferring the atomic population between
the two different internal levels. We assume that all the BEC atoms
initially occupy one of the hyperfine states, let's say $|2\>$. Some
population is then transferred from $|2\>$ to $|1\>$ by means of a
Rabi pulse $\Omega(\rv,t)$. After the pulse, the relative phase
between the BECs in the two levels is proportional to the phase
profile of the Rabi field, as indicated by Eq.~(\ref{gpe}). We may
construct a phase profile for the em field, where the node points of
the field amplitude may correspond to the topological singularities
of the phase of the em field. The em coupling provides then a method
for imprinting these topological singularities on the condensate. In
order to imprint a vortex ring on the $z=0$ plane, centered at the
$z$ axis, we require the em field to exhibit a closed circular phase
singularity in such a way that in the close neighborhood of the
circular node the field is of the form
\beq
\Omega(\rv,t)\simeq {\cal R}(t)\,[ (\rho-\rho_0)+i\gamma z ] = {\cal
R}(t)\, \eta e^{i\theta}\,, \label{ring}
\eeq
where $\rho\equiv(x^2+y^2)^{1/2}$ and $\rho_0$ denotes the radius of
the ring. Here we have also defined $\eta\equiv
[(\rho-\rho_0)^2+\gamma^2 z^2]^{1/2}$ and
$\theta\equiv\arctan{[\gamma z/(\rho-\rho_0)]}$. The field amplitude
(\ref{ring}) vanishes at the ring ($\rho=\rho_0,z=0$) with the
desired $2\pi$ phase winding along any closed loop encircling the
ring, representing one unit of quantized circulation around the
ring. The parameter $\gamma$ describes the anisotropy of the vortex
core. The value $\gamma=1$ corresponds to an isotropic core.

\subsection{Imprinting a vortex ring}

A technique to imprint a vortex ring on the BEC was proposed in
Ref.~\cite{RUO01} by means of constructing an em field amplitude
which exhibits the desired form (\ref{ring}). This was obtained
using
\begin{equation}
\Omega({\bf r})={\Omega_0} \[ \alpha-f(x,y)+i \beta \sin(k z)
\]\,,  \label{ome2}
\end{equation}
Here the Rabi amplitude represents a coherent superposition of a
standing wave along the $z$ axis, a constant field $\alpha$, and the
field $f(x,y)$, which could be a Gaussian field focused weakly on
the $xy$ plane or, alternatively, a superposition of two standing
waves along the $x$ and the $y$ directions:
\begin{align}
f(x,y)&=\exp{\(-{\rho^2\over\xi^2}\) },\label{f1}\\
f(x,y)&={1\over2} \( \cos{2x\over\xi}+\cos {2y\over\xi} \)
\,.\label{f2}
\end{align}
We assume $\rho/ \xi\sim\sqrt{1-\alpha}\ll1$. Then by expanding
Eqs.~(\ref{ome2})-(\ref{f2}) first order in $\rho/ \xi$ and $k
|z|\ll1$, we obtain Eq.~(\ref{ring}) with
$\rho_0=\xi\sqrt{1-\alpha}$ and $\gamma=\beta\xi^2 k/2\rho_0$.
Hence, a constant field with three orthogonal standing waves, or,
alternatively, with a standing wave and a parallel Gaussian beam,
are sufficient to imprint a vortex ring on an atomic BEC. Moreover,
by choosing $\alpha=1-(\beta\xi k/2)^2$, yields an isotropic vortex
core with $\gamma=1$.

The field configuration (\ref{ome2}) is most suitable for microwave
(or longer wavelength) em fields for which the wavelength
$\lambda=2\pi/k$ is longer than the typical radius $R$ of the BEC.
For optical fields $\lambda< R$ the em coupling would create
multiple copies of the ring, displaced from each other by $\lambda$.
However, even for optical fields it should be possible to shape the
wave fronts of the coupling lasers in order to avoid rapid phase
variation at the length scale $\lambda$. This could be done, e.g.,
by using microlens arrays \cite{DUM02}, laser beams co-propagating
in the direction of the standing wave $z$, or beams with the wave
vectors nearly perpendicular to $z$. Here we show that another
alternative is to consider two-photon transitions via some
intermediate level $|3\>$ and use slightly different wave numbers
for the lasers. Consider the first transition $|2\>\rightarrow |3\>$
to be induced by four co-propagating laser beams, one of which is
weakly focused on the $xy$ plane,
\begin{align}
\Omega_2(\rv)= &{\bar\Omega_2}\[ \( \alpha-e^{-\rho^2/\xi^2} \)
e^{ikz} \right. \nonumber\\ &\left. + {\beta\over2} \(
e^{i(k+k')z}-e^{i(k-k')z} \) \] \,. \label{emring1}
\end{align}
Here we assume that $k'\ll k$ and the coupling frequency
$\Omega_2=\dv_{23}\cdot\Ev_2^+/\hbar$ is in this case determined in
terms of the atomic dipole matrix element $\dv_{23}$ of the
transition $|2\>\rightarrow |3\>$ and the positive frequency
component of the driving electric field $\Ev_2^+$. The second
transition $|3\>\rightarrow |1\>$ is driven by a non-focused
co-propagating field $\Omega_1(\rv)=\bar\Omega_1\exp{(ikz)}$, which
cancels the rapid phase variation of $\Omega_2$ along the $z$ axis.
In the limit of large detuning $\Delta$ of the laser $\Omega_2(\rv)$
from the resonance of the $|2\>\rightarrow |3\>$ transition, the
effective Rabi frequency for the $|2\>\rightarrow |1\>$ transition
then reads
\begin{align}
\Omega_{21}(\rv)&\simeq
{2\Omega_2(\rv)\Omega^*_1(\rv)\over\Delta}\nonumber\\&=
{2\bar\Omega_2\bar\Omega_1^*\over\Delta}  \[
\alpha-e^{-\rho^2/\xi^2} +i \beta \sin{k'z}\]\,.
\end{align}
This has the desired form (\ref{ome2}). In addition, even though we
may have $k R\gg 1$, it should be possible to choose $k'\ll k$, so
that $k' R\alt1$. This allows us to create precisely one vortex ring
in the BEC.

In a more challenging scheme, one may construct an optical lattice
with a long periodicity \cite{DUM98} by means of superposing two
separate two-photon transitions
\beq
\sin[(k_1-k_2) z]= {1\over 2i} \( e^{ik_1z}e^{-ik_2 z}- e^{-ik_1
z}e^{ik_2 z}\)\,,
\eeq
such that $(k_1-k_2)R\alt 1$. Here each exponent factor represents a
running wave transition by one photon.

In the following, we demonstrate how the vortex ring could also be
imprinted using only two Gaussian laser beams providing an
experimentally simple scheme for a controlled creation of vortex
rings.

\subsection{Imprinting a vortex ring using Gaussian beams}

Here we show that a vortex ring defect can also be imprinted on an
atomic BEC in a controlled way by means of two parallel Gaussian
laser beams. The advantage of this scheme, as compared, e.g., to
Eq.~(\ref{emring1}) is the obvious simplicity of the field
configuration and the common experimental availability of the beams.
The complete expression of the Gaussian laser beam reads:
\begin{align}
{\cal G}(\rv,w_0) & =   {w_0\over w(z)} \exp{\[ i kz
-i\arctan{z\over
z_0} + ik{\rho^2\over 2R(z)}\] }\nonumber\\
& \times \exp{\[ -{\rho^2\over w(z)^2}\] }\,, \label{gauss}
\end{align}
where $w(z)\equiv w_0 \sqrt{1+z^2/z_0^2}$ and $R(z)\equiv
z(1+z_0^2/z^2)$. Here $w_0$ represents the minimum beam waist and
$z_0\equiv \pi w_0^2/\lambda$ is the Rayleigh range, the distance
over which the focusing is increased to $\sqrt{2}w_0$ due to
diffraction. In a weakly focused beam, for $\rho\ll w_0$, $|z|\ll
z_0$, ${\cal G}(\rv,w_0)\simeq\exp{(ikz-\rho^2/w_0^2)}$.

We take the two Gaussian beams as a phase-coherent superposition:
\beq
\Omega_r(\rv)= \Omega_0\, [{\cal G}(\rv,w_0)+c{\cal G}
(\rv,w_0')]\,, \label{gaussring}
\eeq
for some complex $c$. It is straightforward to see that by choosing
\beq
c= -\exp{\[ - \rho_0^2\( {1\over w_0^2}- {1\over w_0'^2} \) \] }\,,
\label{c}
\eeq
the field configuration (\ref{gaussring}) exhibits a circular node
at ($\rho=\rho_0,z=0$) with the $2\pi$ phase winding around the
ring. The complete destructive interference of the two Gaussian
beams does not occur outside the plane of the minimum beam focus
size due to diffraction. By modifying the relative amplitude and the
focusing of the beams, the size of the ring can be changed in a
controlled way. The anisotropy of the core is determined by
\beq
\gamma= {w_0^2 w_0'^2-(w_0^2+w_0'^2)\rho_0^2\over w_0^2 w_0'^2
\rho_0 k}\,.
\eeq

In order to avoid the rapid phase variation of the Gaussian laser
along the $z$ axis to be imprinted on the BEC, it is again
advantageous to consider a two-photon transition via some
intermediate level $|3\>$ using co-propagating lasers, so that,
e.g., the field $\Omega_r(\rv)$ in Eq.~(\ref{gaussring}) drives the
transition $|2\>\rightarrow |3\>$ and a non-focused co-propagating
field $\Omega_{31} \exp{(ikz)}$ drives the transition
$|3\>\rightarrow |1\>$. In the limit of large detuning $\Delta$, the
effective Rabi frequency for the $|2\>\rightarrow |1\>$ transition
then reads $2\Omega_{31} \Omega_r(\rv)\exp{(-ikz)}/\Delta$. The
corresponding spatially non-uniform laser-induced level shifts are
$\delta_1(\rv)\simeq 2|\Omega_{31}(\rv)|^2/\Delta+\delta_{12}$ and
$\delta_2(\rv)\simeq 2 |\Omega_{r}(\rv)|^2/\Delta$, where
$\delta_{12}$ is the effective two-photon detuning from the
resonance of the internal transition $|2\>\rightarrow|1\>$.

\begin{figure}\vspace{-0.5cm}
\includegraphics[width=0.5\columnwidth]{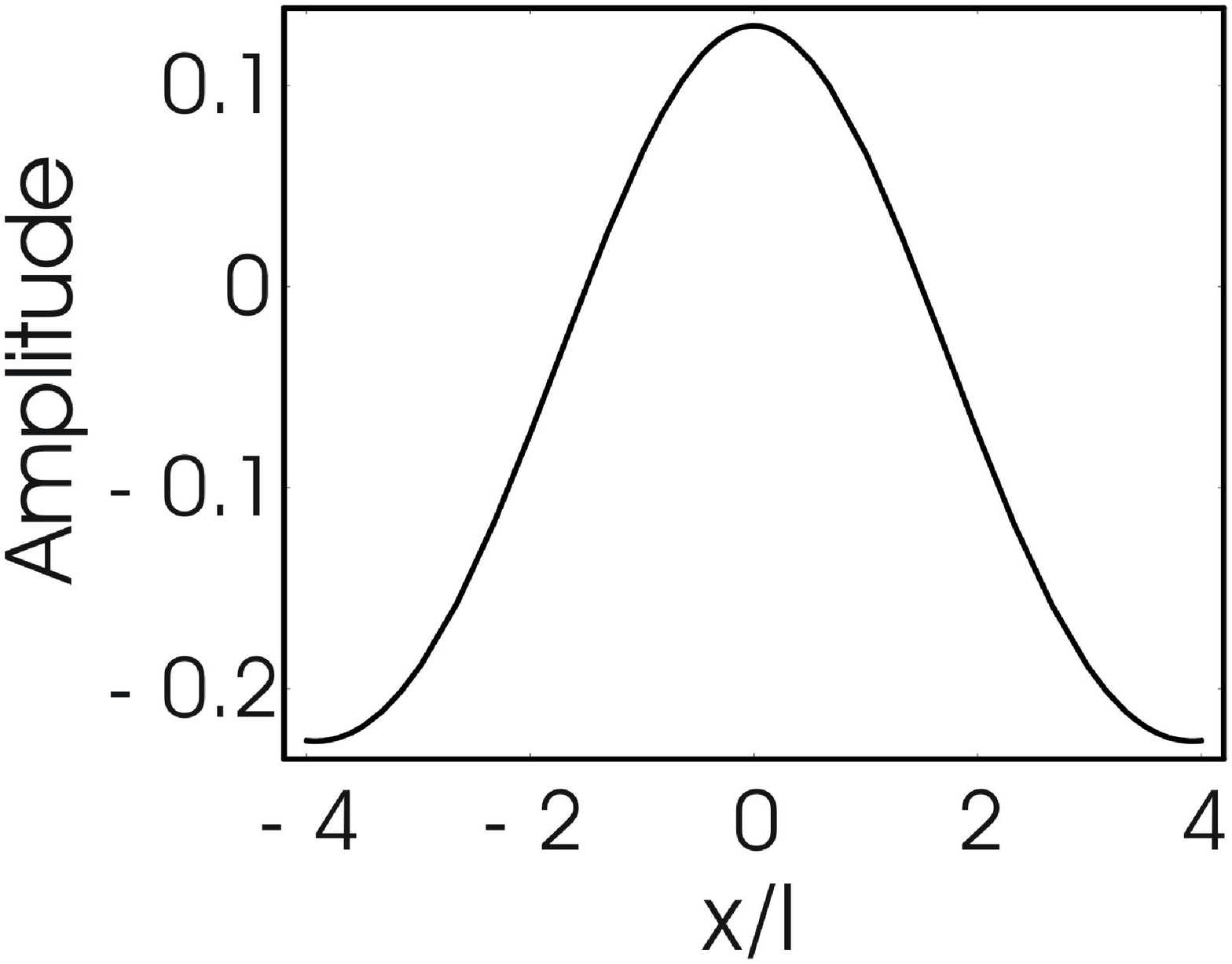}\hspace{0.1cm}
\includegraphics[width=0.44\columnwidth]{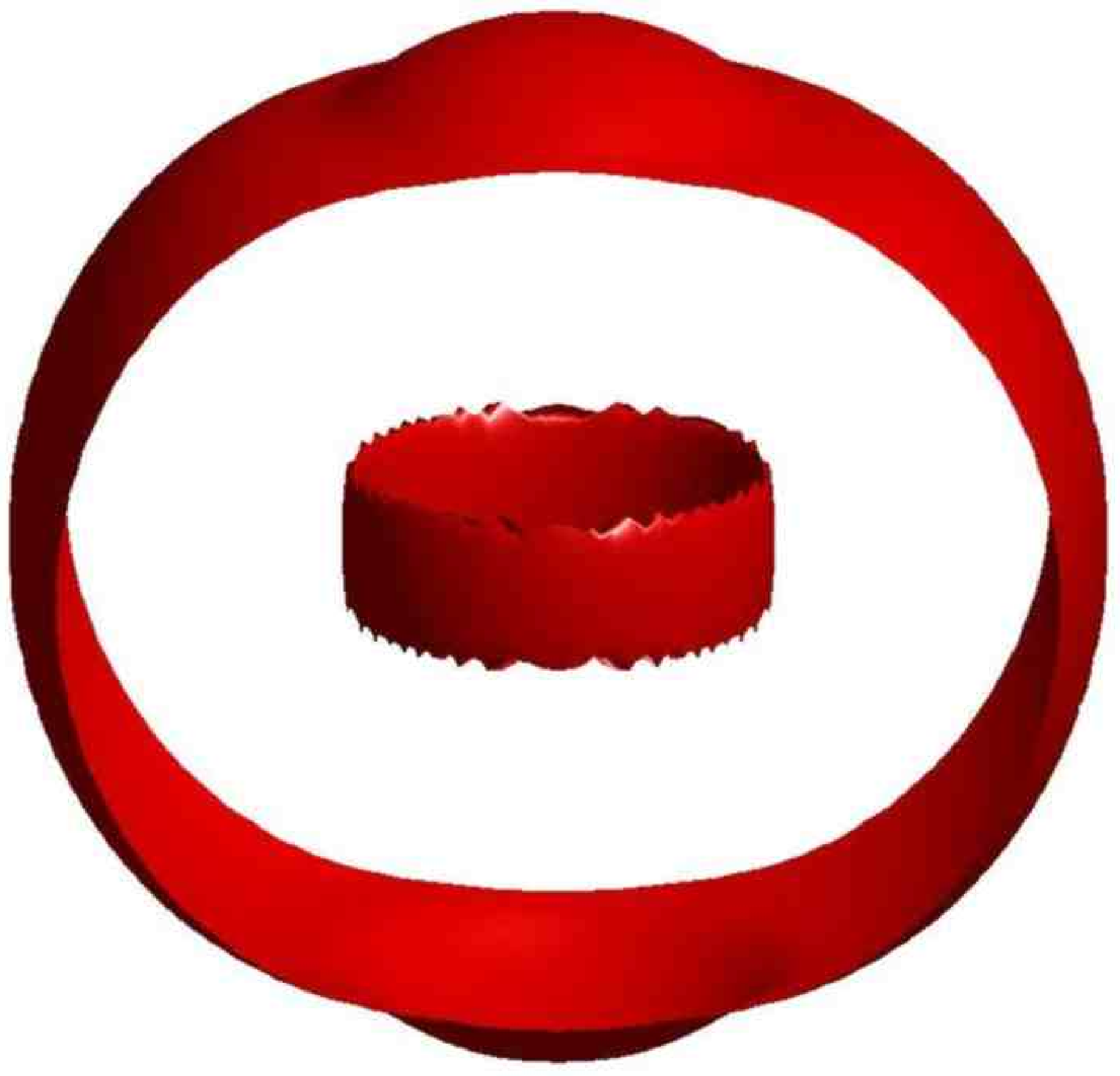}
\vspace{-0.3cm} \caption{The amplitude of the superposition of two
Gaussian laser beams at $z=0$ (on the left). The cylindrically
symmetric amplitude exhibits a node point at $\rho_0=1.5l$. The
constant surface plot of a vortex ring in an atomic BEC (on the
right) with the phase profile imprinted from the two Gaussian laser
beams. The Rabi frequency $\propto e^{-i k_2 z}\Omega_r$. Here the
beam parameters are $w_0=3.0l$, $w_0'=4.5l$, $kl=7$, $k_2l=6.86$,
and the radius of the ring $\rho_0=1.5l$. As a result of the
imprinted phase pattern, the vortex ring slowly moves along the beam
axis.} \label{gaussfig}
\end{figure}

In Fig.~\ref{gaussfig} we show an example of a vortex ring in a
two-component BEC, obtained by imprinting the phase profile of the
two Gaussian laser beams from Eq.~(\ref{gaussring}). We use the
parameters of $^{87}$Rb with the nonlinearity $\kappa_{21}=500\hbar
\omega l^3$ and the radius of the ring $\rho_0=1.5l$, where
$l=\sqrt{\hbar/m\omega}$. The ring occupies the level $|1\>$ with
the atom numbers $N_2/N_1\simeq6.2$. For the chosen parameters, the
vortex core is noticeably anisotropic with $\gamma\simeq 0.06$. A
more isotropic core may be obtained, e.g., by decreasing the radius
of the ring by means of changing the value $c$ according to
Eq.~(\ref{c}). For $\rho_0=1.0l$ we obtain $\gamma\simeq 0.12$ and
for $\rho_0=0.5l$, $\gamma\simeq 0.27$. As a result of the nonlinear
evolution and dissipation, an anisotropic vortex core in an atomic
BEC is generally expected to relax towards an isotropic core shape.

Experimentally, one could prepare the phase-coherent field
configuration (\ref{gaussring}) using a single laser source which is
split in two and later recombined after appropriately modifying the
relative amplitude and the beam focusing. One possible experimental
limitation might then be a slight misalignment of the two recombined
Gaussian beams. We also tried field configurations with the
propagating axes of the two beams displaced. It is quite easy to
show that the vortex ring singularity Eq.~(\ref{gaussring}) is
robust against perturbations where the two Gaussian beams are
slightly modified.

\subsection{Multi-beam superpositions}

Two beam configurations, such as Eq.~(\ref{gaussring}), or more
complicated multi-beam superpositions can also be prepared using
diffractive optical components. In particular, computer-generated
holograms and spatial light modulators may be used to prepare the
desired optical field superpositions which are required to imprint
the vortices on the atomic BECs. By using a third Gaussian beam in
Eq.~(\ref{gaussring}) we may better optimize the field parameters
according to a steady-state vortex ring solution, or, alternatively,
to engineer some desired excitations in the ring. With the
superposition: \beq \Omega_r(\rv)= \Omega_0\, [{\cal
G}(\rv,w_0)+c_1{\cal G} (\rv,w_1)+c_2{\cal G} (\rv,w_2)]\,,
\label{gaussringx} \eeq we may, e.g., choose the parameters $c_1$
and $c_2$ according to the given vortex ring radius and the core
isotropy, while $w_0$, $w_1$, and $w_2$ may be optimized to
determine the spatial profile of the BEC wave function.

\subsection{Knotted vortex lines}

More complicated field superpositions than the one shown in
Eq.~(\ref{gaussringx}) have been successfully created using spatial
light modulators. In Refs.~\cite{leach04,leach04b} a field
configuration of four optical Laguerre-Gaussian beams, with each
exhibiting zero, one, or two units of orbital angular momentum, were
generated using a liquid crystal array acting as a phase mask and
controlling both the field amplitude and the phase. Such a field
superposition possesses a phase singularity forming a closed knotted
loop \cite{BER01,BER01b} (a torus knot) in a paraxial field and, in
principle, could also be imprinted inside a BEC, provided that the
hologram has a sufficient spatial resolution in order to allow the
beams to be focused inside the atomic cloud.

\section{ Engineering vortex reconnections}
\label{reconnections}

\subsection{Preparing multiple vortex rings}

We may use the techniques described in the previous section also to
imprint multiple vortex rings on the BEC. Generally this can be done
by using multi-photon transitions, where the effective Rabi pulse is
of the form:
\beq
\Omega(\rv,t)\simeq {\cal R}(t)\, \prod_{j=1}^n [
(\rho^{(j)}-\rho^{(j)}_0)+i(z^{(j)}-z_0^{(j)}) ]^{p_j}\,,
\label{multiring}
\eeq
where we write
$\rho^{(j)}=[(x^{(j)}-x_0^{(j)})^2+(y^{(j)}-y_0^{(j)})^2]^{1/2}$,
representing $n$ vortex rings on the plane $z^{(j)}=0$, centered at
($x^{(j)}=x_0^{(j)},y^{(j)}=y_0^{(j)},z^{(j)}=z_0^{(j)}$), with the
radius given by $\rho^{(j)}_0$. All the vortex rings in
Eq.~(\ref{multiring}) do not need to have the same orientation, but
the axis $\rv^{(j)}$ may involve different spatial rotations with
respect to some fixed axis $\rv$. The exponent $p_j$ in
Eq.~(\ref{multiring}) denotes the topological charge of the vortex
ring $j$.

\begin{figure}[!b]\vspace{-0.5cm}
\includegraphics[width=0.49\columnwidth]{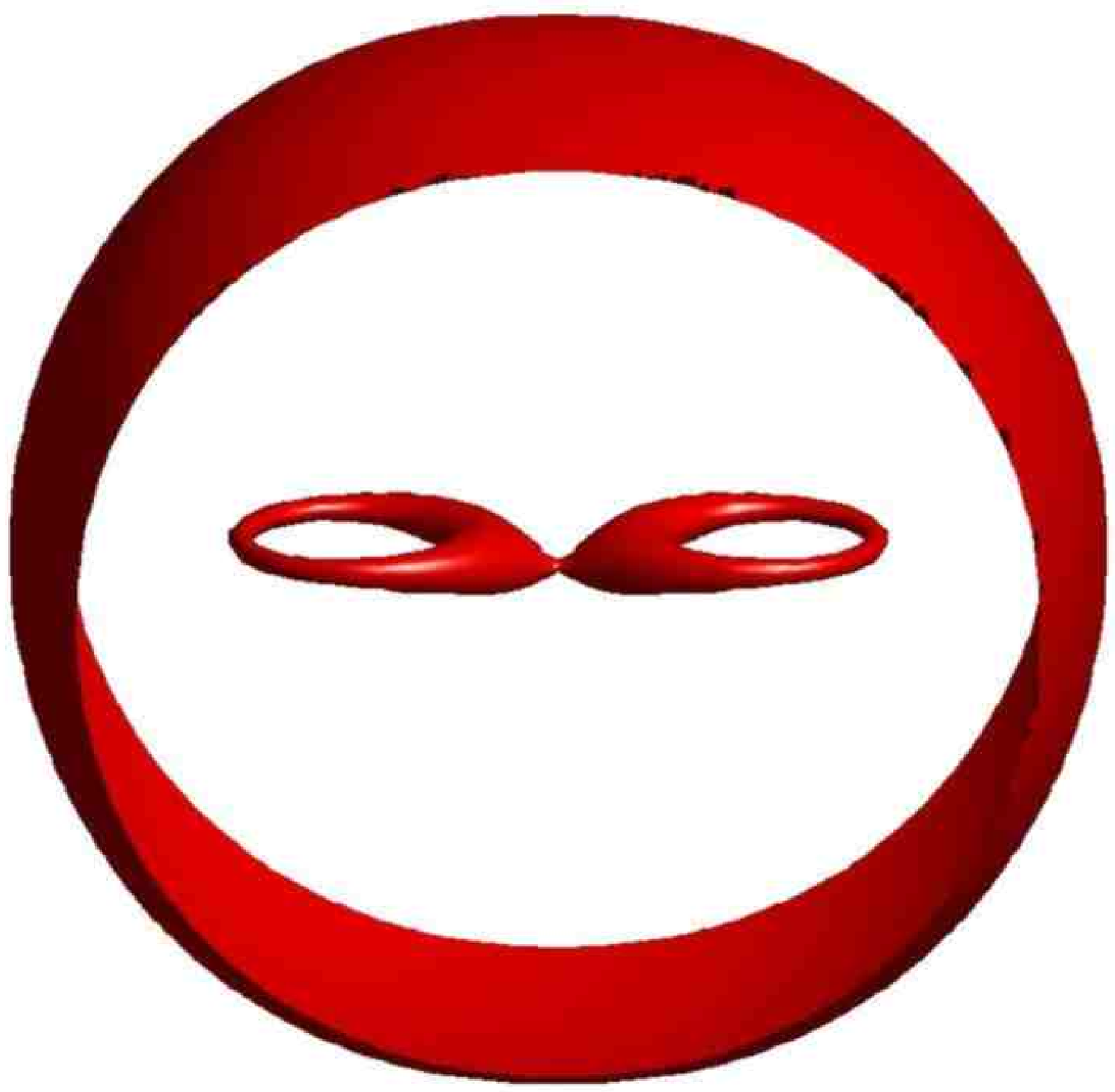}
\includegraphics[width=0.49\columnwidth]{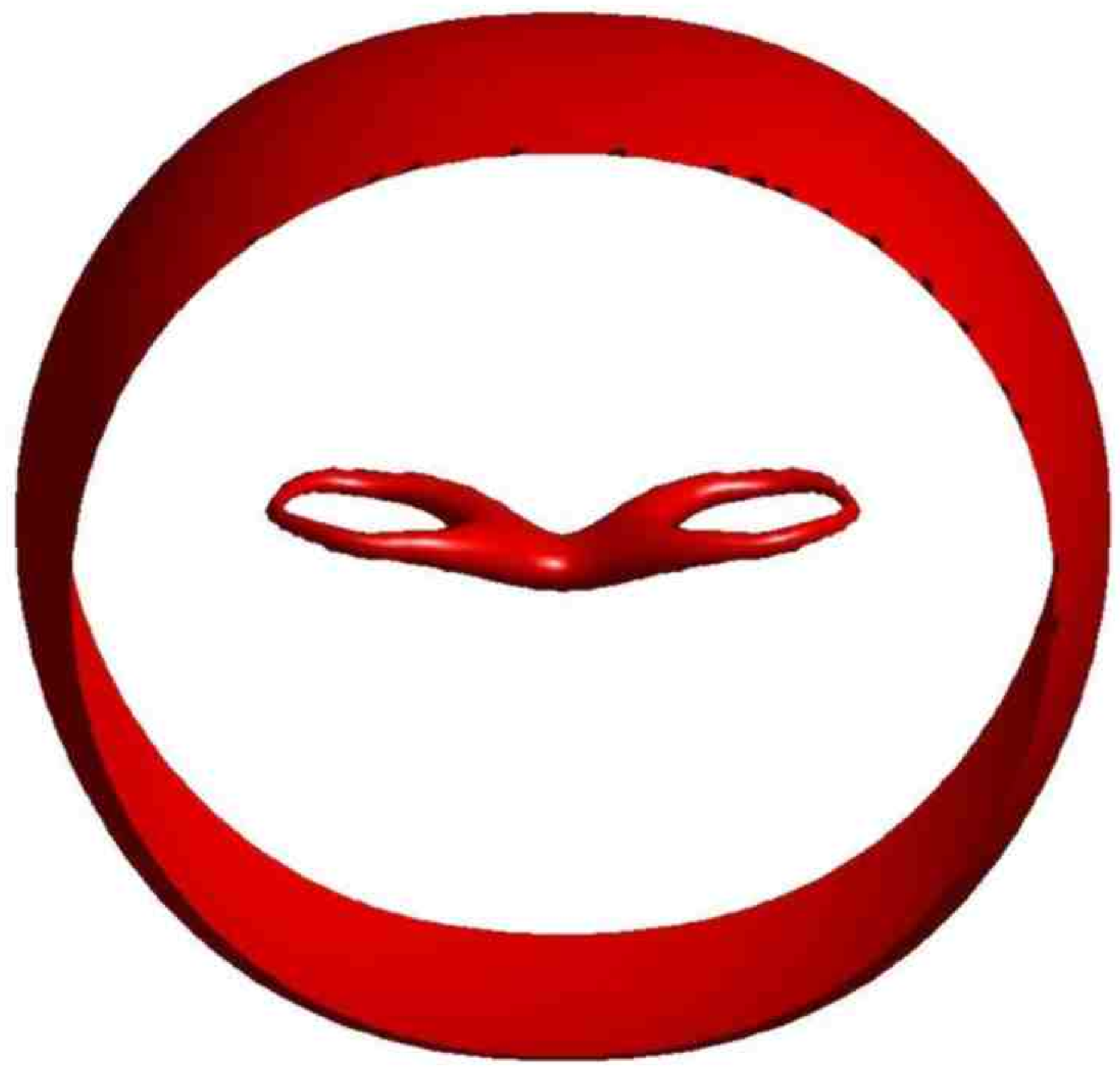}
\includegraphics[width= 0.49\columnwidth]{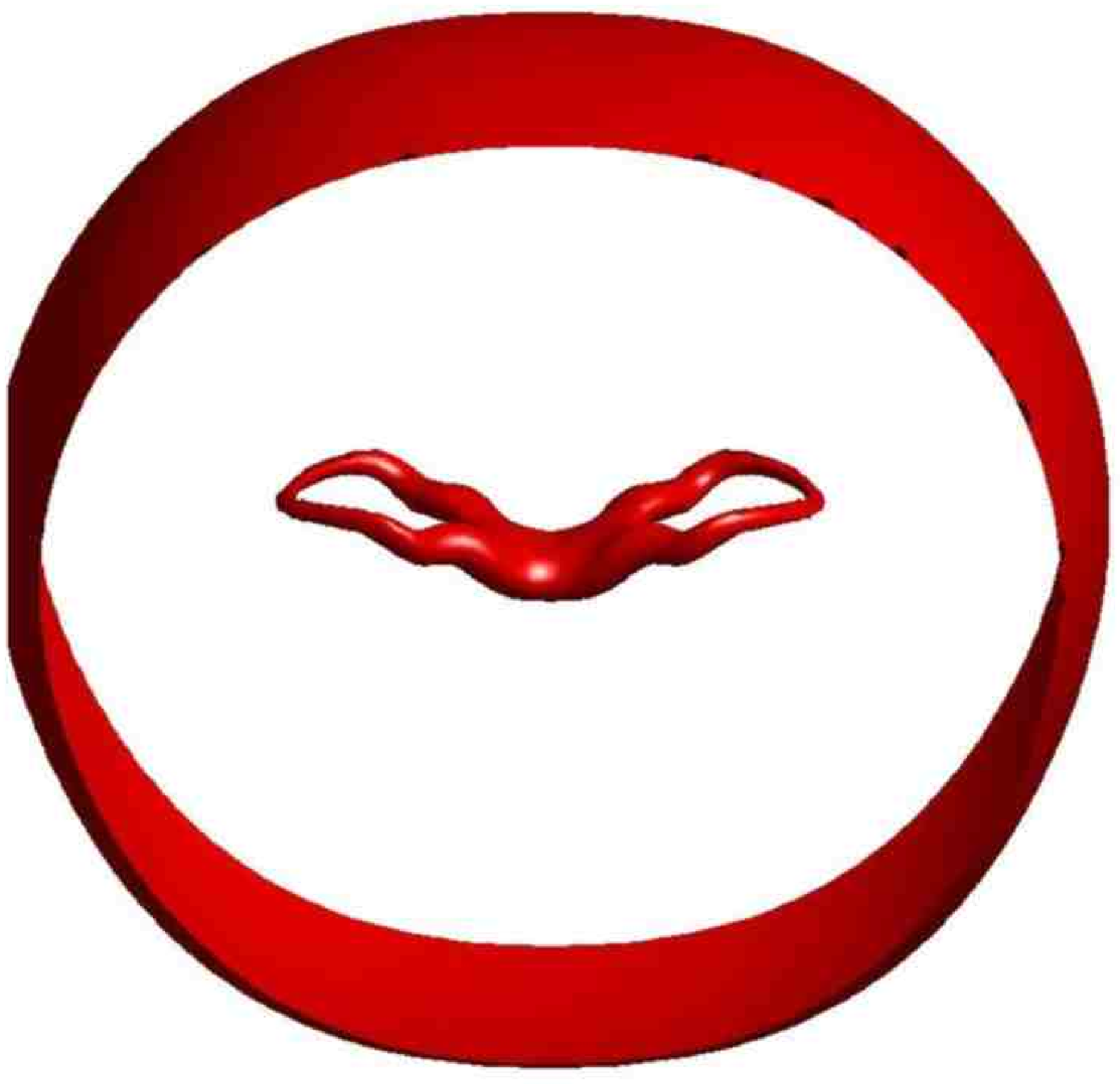}
\includegraphics[width=0.49\columnwidth]{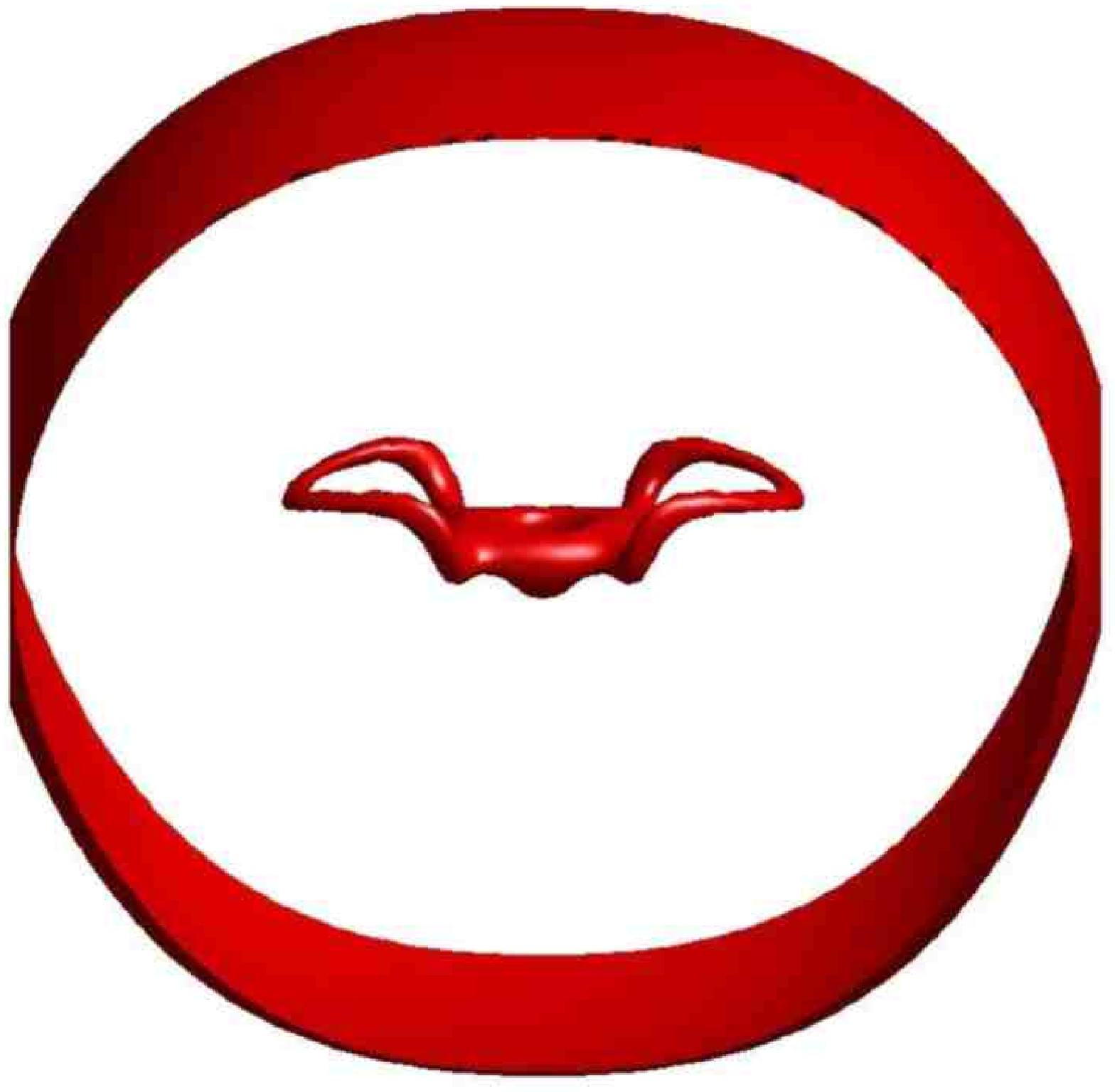}
\includegraphics[width=0.49\columnwidth]{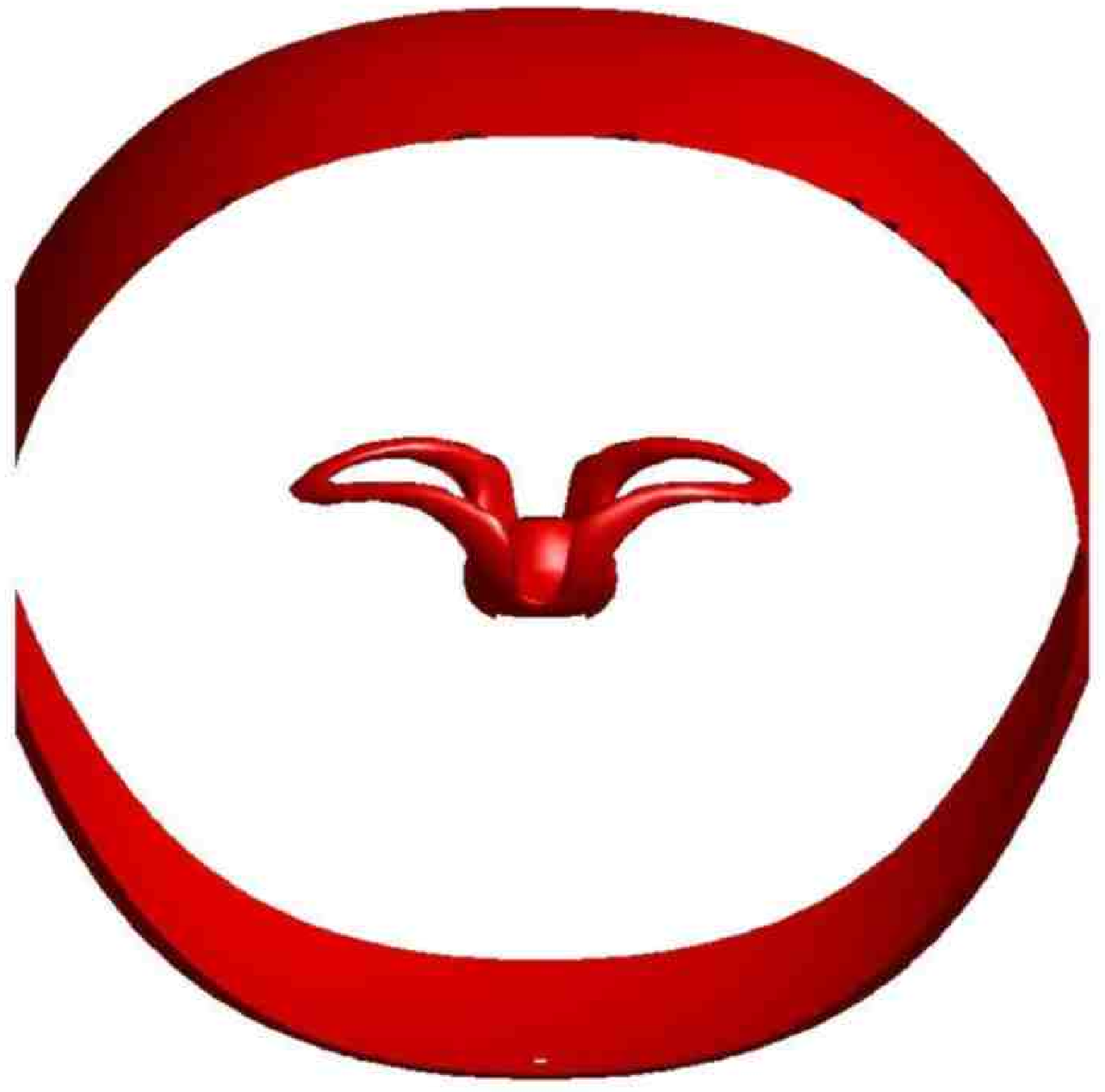}
\includegraphics[width= 0.49\columnwidth]{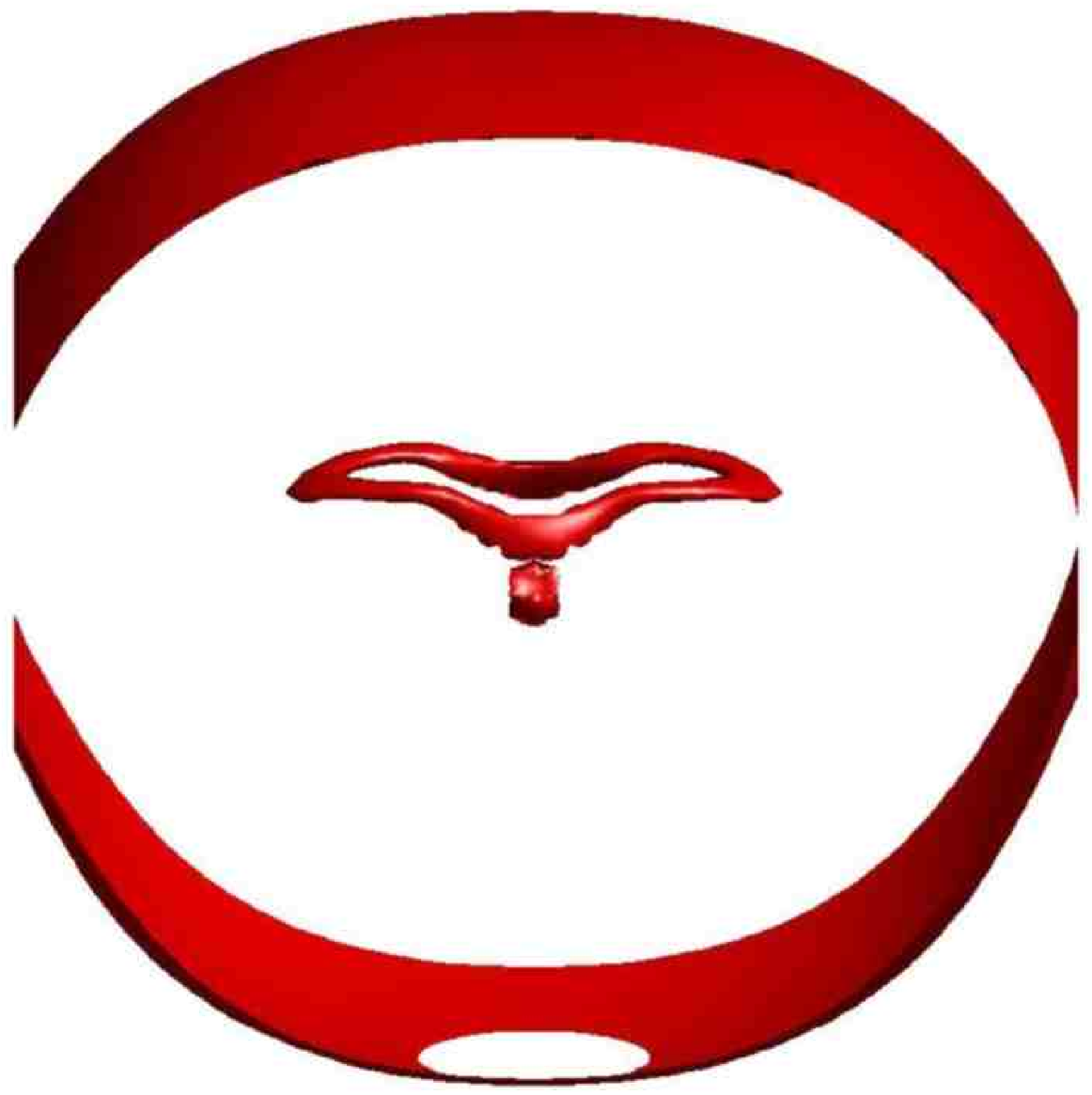}
\includegraphics[width=0.49\columnwidth]{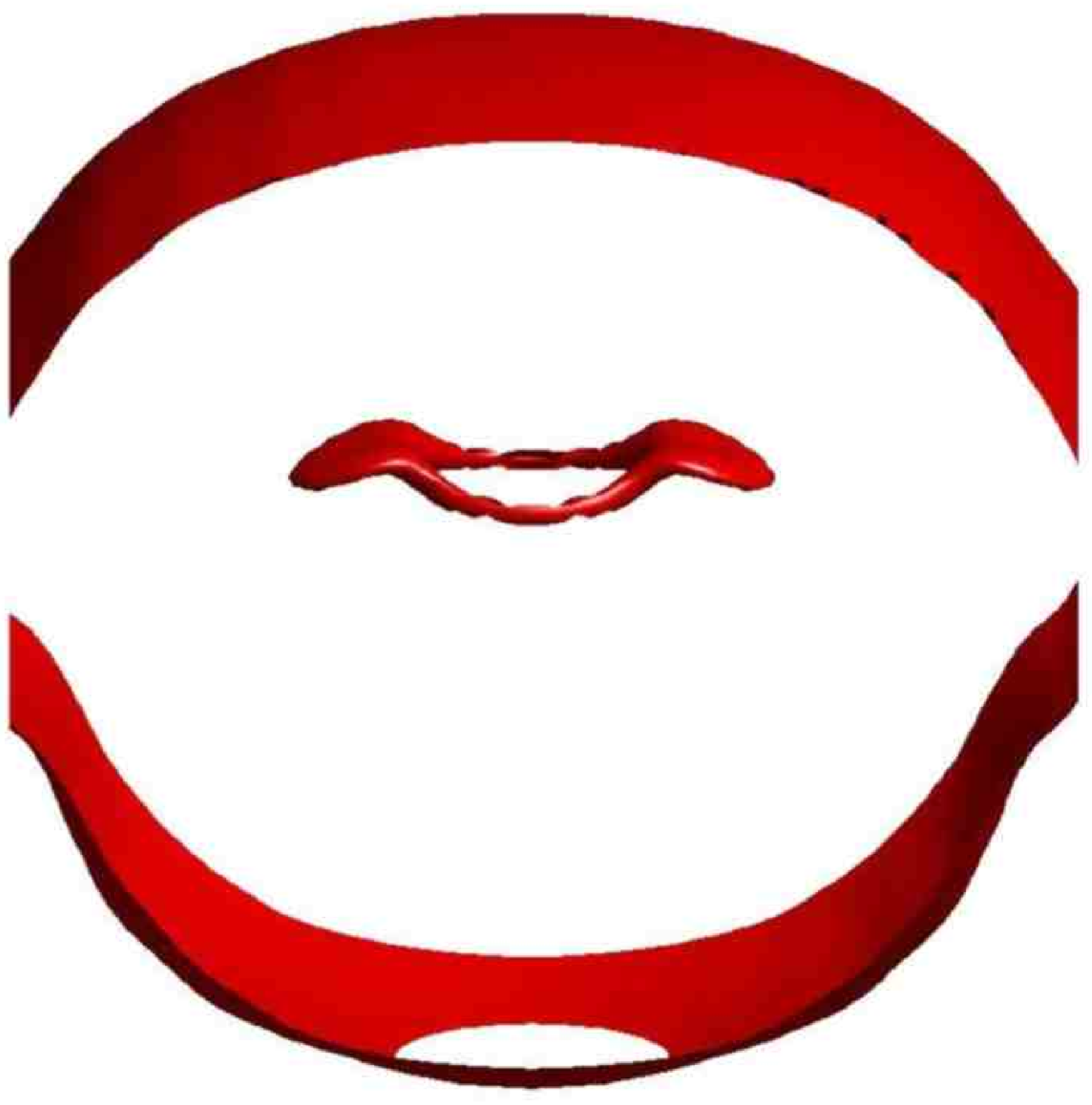}
\includegraphics[width=0.49\columnwidth]{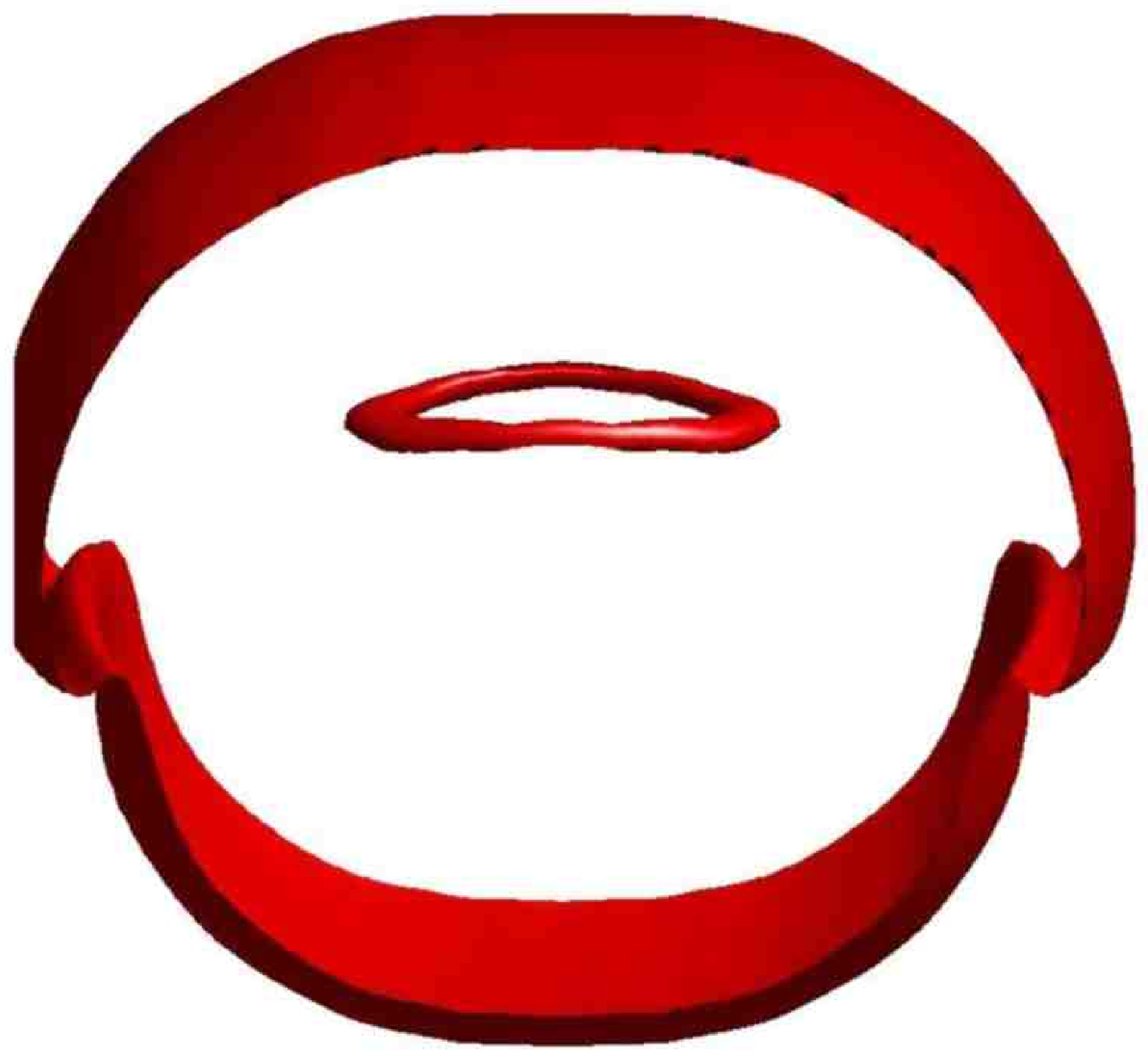}
\vspace{-0.3cm} \caption{The snapshot images of the isosurface plots
during the time evolution of two merging vortex rings in a trapped
two-component BEC. We show the BEC component containing the two
vortex rings, while the other component exhibits a uniform phase
profile. The initial state was prepared by means of an em coupling
field between the two BEC components, which imprints the desired
phase singularities on the matter field while changing the internal
level of the atoms. In order to relax the energy of the initial
state, we integrated the wave functions in imaginary time for the
duration of $t\omega=0.06$ before the real time evolution. We show
the isosurface plots at times
$t\omega=0.1,0.2,0.3,0.4,0.5,0.6,1.0,1.4$.} \label{ring1}
\end{figure}

We numerically studied the preparation of two nonoverlapping vortex
rings by integrating the coupled GPE (\ref{gpe}) in the presence of
the em coupling. In Fig.~\ref{ring1} we show an example of two
vortex rings with the radii $\rho_0=0.9l$ prepared on the $z=0$
plane.  Both rings are displaced from the trap center in opposite
directions by $1.5l$. Here the vortex rings are created by means of
the em field $\Omega(\rv-\rv_1)\Omega(\rv-\rv_2)$ (for
$\rv_1=-\rv_2=1.5l \hat{y}$) with $\Omega(\rv)$ determined by
Eqs.~(\ref{ome2}) and~(\ref{f1}). The wavelength $\lambda=26l$,
$\beta\simeq 8.3\times 10^{-3}$, and the width $\xi=30l$. We use the
parameters of $^{87}$Rb with the atoms initially occupying the level
$|2\>$. The nonlinearity is $\kappa_{21}=430\hbar \omega l^3$. A
short pulse $t\omega=0.1$ is applied to transfer population to the
level $|1\>$ in order to prepare the vortex rings. This could be,
e.g., a two-photon transition via some intermediate level $|3\>$,
where the field $\Omega(\rv-\rv_1)$ drives the transition
$|2\>\rightarrow |3\>$ and the field $\Omega^*(\rv-\rv_2)$ drives
the transition $|3\>\rightarrow |1\>$. Alternatively, we could have
used the Gaussian laser beams (\ref{gaussring}), e.g., driving a
two-photon $\Lambda$ three-level transition, where the beams
inducing the second transition are co-propagating with the first
pair of beams, but are multiplied by the phase factor $(-1)$ and
displaced in space by the amount of $3l$. By modifying the strength
and the duration of the Rabi pulse we can control the excitation of
the vortex ring pair. This allows us to study the dynamics of the
vortex ring interactions for different initial conditions.

\subsection{Vortex ring reconnections}

Vortex reconnections represent basic vortex-vortex interactions.
These are of fundamental interest in their own right, as well as
providing a useful model to investigate superfluid turbulence which
can be considered as being due to the dynamics of a dense tangle of
quantized vortex filaments \cite{DON91}. The evolution of such a
tangle crucially depends on the vortex reconnections. The vortex
reconnections also have an analogy with the models describing the
evolution of cosmic strings in the early Universe cosmology
\cite{BAU96}. At very low temperatures, with negligible normal fluid
fraction, the superfluid turbulence and the vorticity is expected to
decay as result of sound emission \cite{VIN00}. Sound emission may
occur due to vortex motion or reconnection.

The proposed techniques of vortex imprinting provide a useful tool
in investigating vortex reconnections in superfluids. Atomic BECs
can be cooled down to very low temperatures with negligible normal
fluid component. By modifying the pulse parameters of the imprinting
process we may also control the level of excitation in the system.
Moreover, vortices in two-component BECs offer several additional
advantages.  First, it is possible to monitor the system in real
time: The filling of the vortex cores by the other BEC significantly
increases the core size, as compared to vortices in a
single-component BEC, and makes them observable {\it in situ}, even
without a ballistic expansion.  Second, the nonlinearity of the
vortex excitations can be easily controlled by varying the relative
density in the two components (via the strength and the duration of
the coupling laser fields).

The vortex reconnections have previously been numerically studied
using the GPE for the parallel and crossed line vortices
\cite{KOP93,berloff} and for colliding vortex rings
\cite{KOP96,adams} in a homogeneous single-component BEC. In
Refs.~\cite{KOP96,adams} two vortex rings with an initial relative
velocity were scattered from each other, varying the angle of
intersection and the impact parameter. As a result, a substantial
loss of vortex line length was observed, which was attributed to
sound emission.

As an application of the multi-ring imprinting techniques, we showed
here how to create two initially static vortex rings in a trapped
inhomogeneous two-component BEC, which will later interact and
collide.  The two vortex rings on the same $z=0$ plane were
imprinted using em fields, as explained in the previous section. Due
to the filling of the vortex core by the second BEC component, an
isolated vortex ring would approximately remain in the same location
in the trap on the time scale of the trap period. If the initial
separation of the rings is sufficiently small, the vortex rings are
attracted to each other and merge locally, after the em coupling
fields have been turned off.

Qualitatively, the long-range interaction energy between vortex
filaments $s$ and $s'$ with the vorticity $\kappa$ and $\kappa'$ may
be obtained, analogously to the magnetostatics, from \cite{LAM45}
\beq
E= {n\kappa\kappa'\over 8\pi} \int  {d{\bf s}\cdot d{\bf s}'\over
|\rv(s)-\rv(s')|}\,. \label{vorene}
\eeq
Two antiparallel vortex filaments are attracted to each other. Once
the vortex cores are close enough, they merge at the point of the
closest approach. Since the facing segments of the two similarly
circulating rings on the same plane represent vorticity with the
opposite circulations, these segments are attracted to each other
and eventually they annihilate. Consequently, the two rings are
joined together. By integrating Eq.~(\ref{vorene}) with an
appropriate cut-off, we may also obtain the energy of an isolated
vortex ring as $E\propto n \rho_0 [\ln (\epsilon \rho_0)-2]$, where
$n$ denotes the density and the parameter $\epsilon$ depends, e.g.,
on the core thickness \cite{ROB71,berloff} and where $\rho_0$
denotes the ring radius. Close to equilibrium the reduction of
vortex core length can indicate the emitted sound energy
\cite{lead03}.

In the numerical studies of the two-component BEC dynamics, we use
initial states corresponding to different strength of excitation.
These could be created by changing the strength and the duration of
the em pulse in the imprinting process. However, for computational
simplicity, here we always use the same pulse parameters, but relax
the initial state by means of varying length of imaginary time
evolution before the actual dynamics.

\begin{figure}[!b]\vspace{-0.5cm}
\includegraphics[width=0.49\columnwidth]{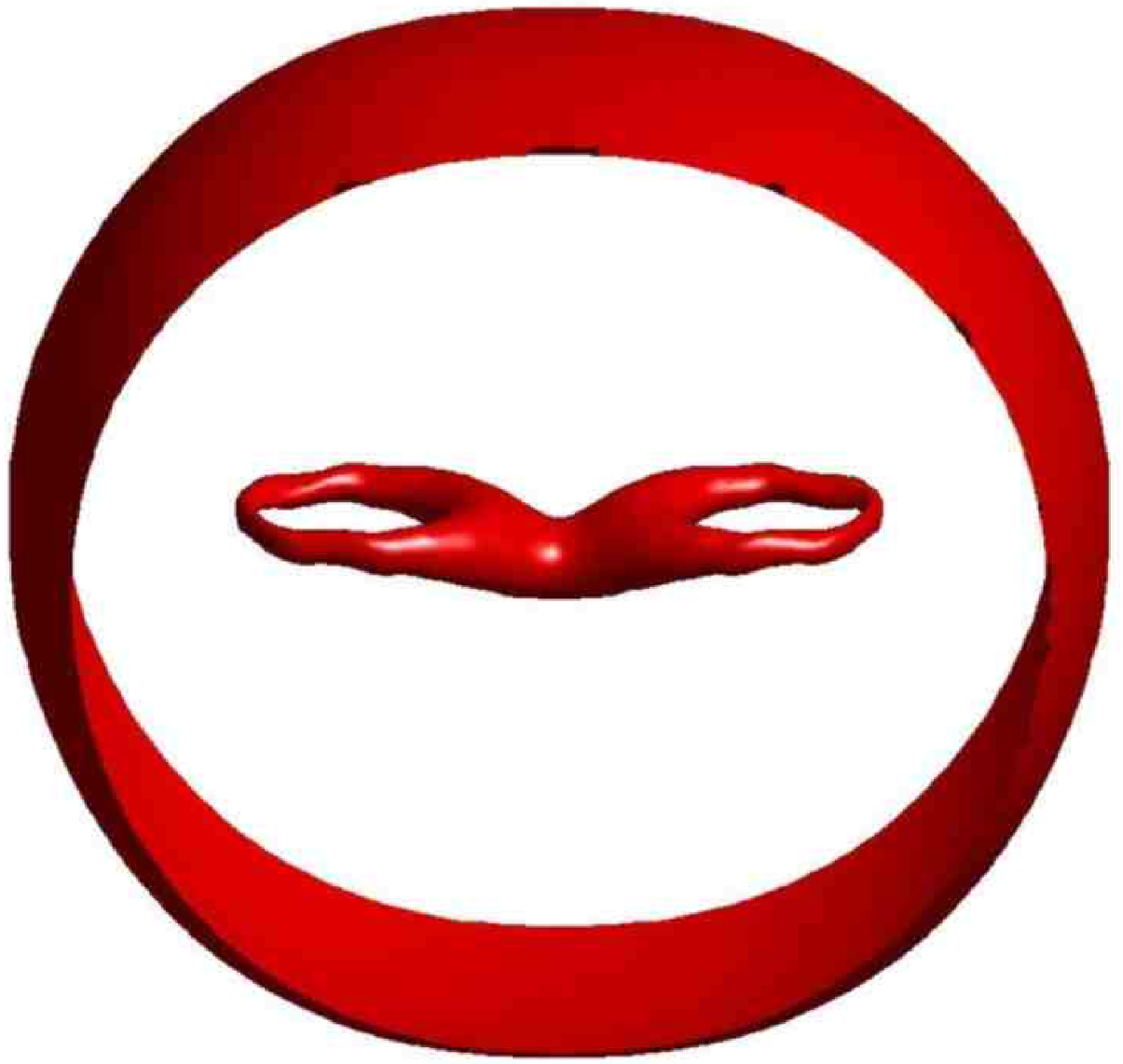}
\includegraphics[width=0.49\columnwidth]{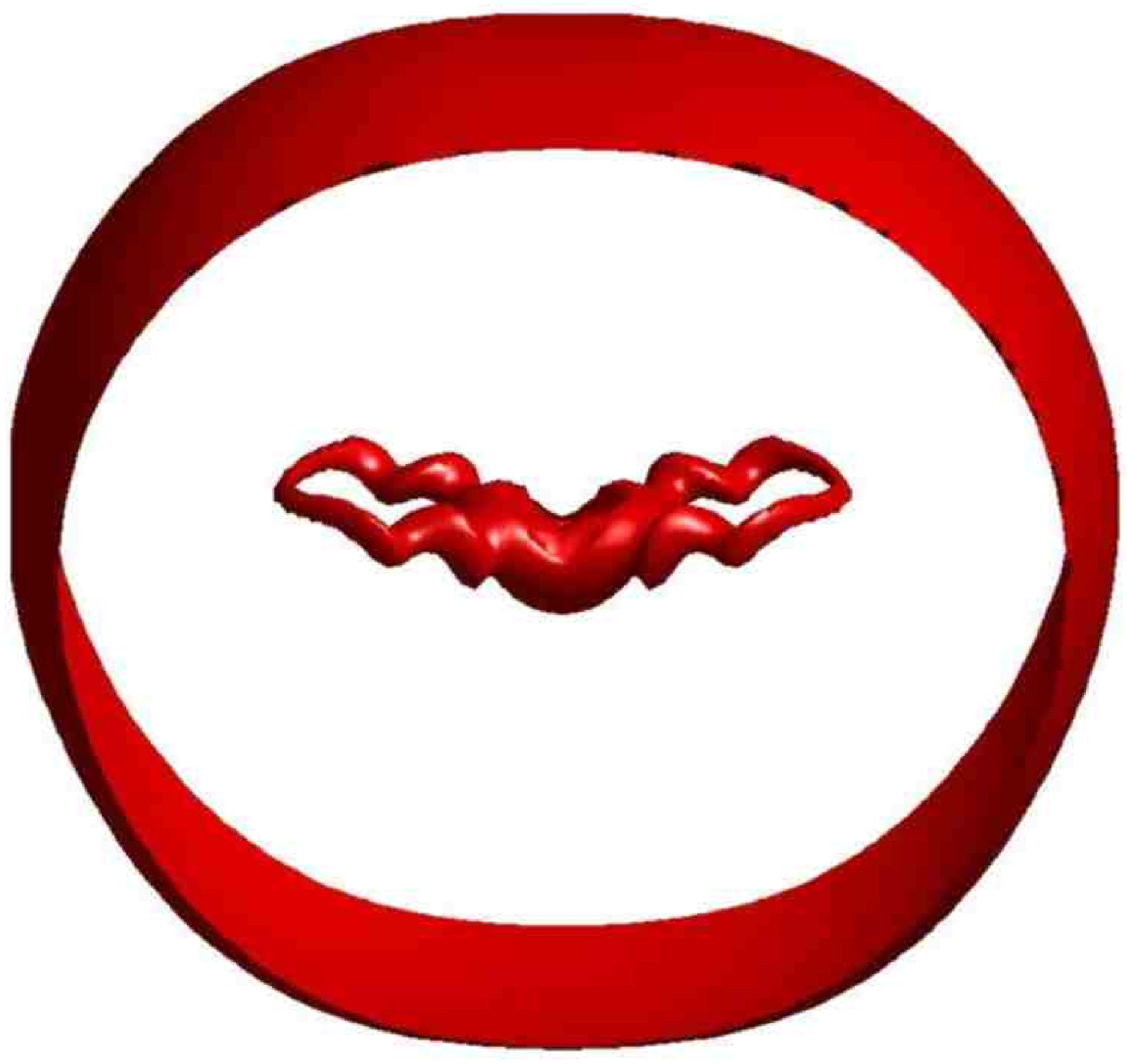}
\includegraphics[width= 0.49\columnwidth]{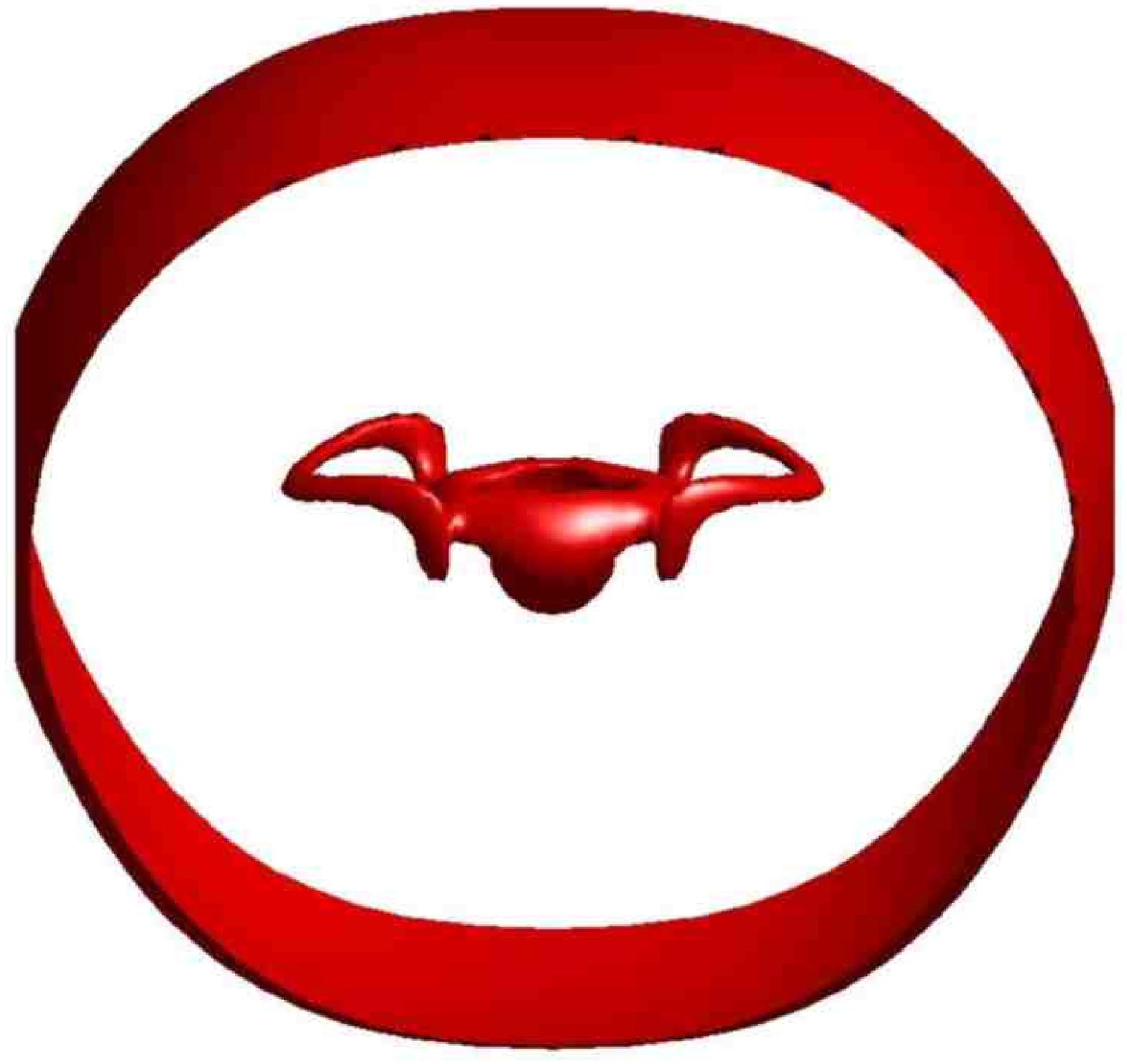}
\includegraphics[width=0.49\columnwidth]{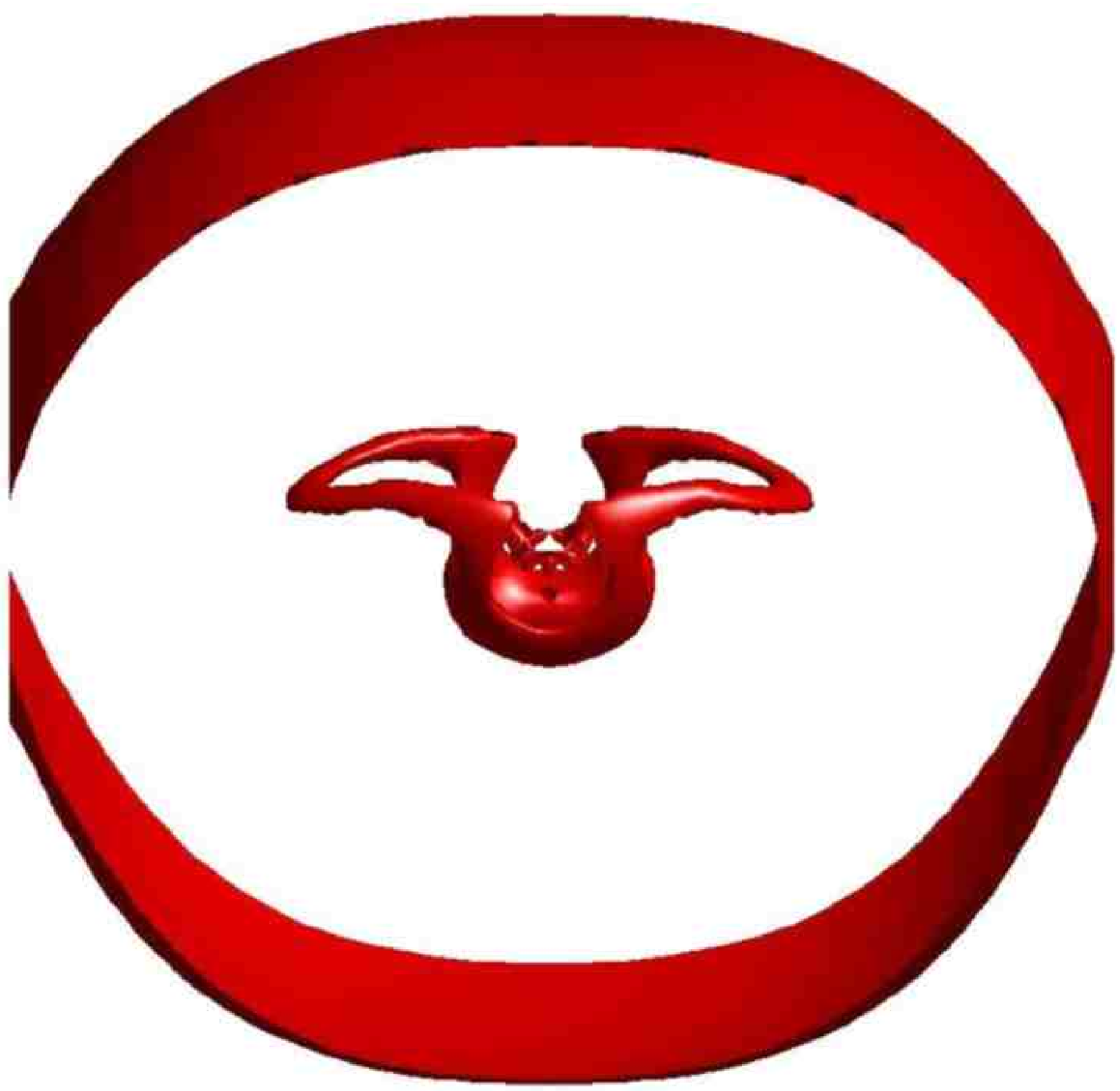}
\includegraphics[width=0.49\columnwidth]{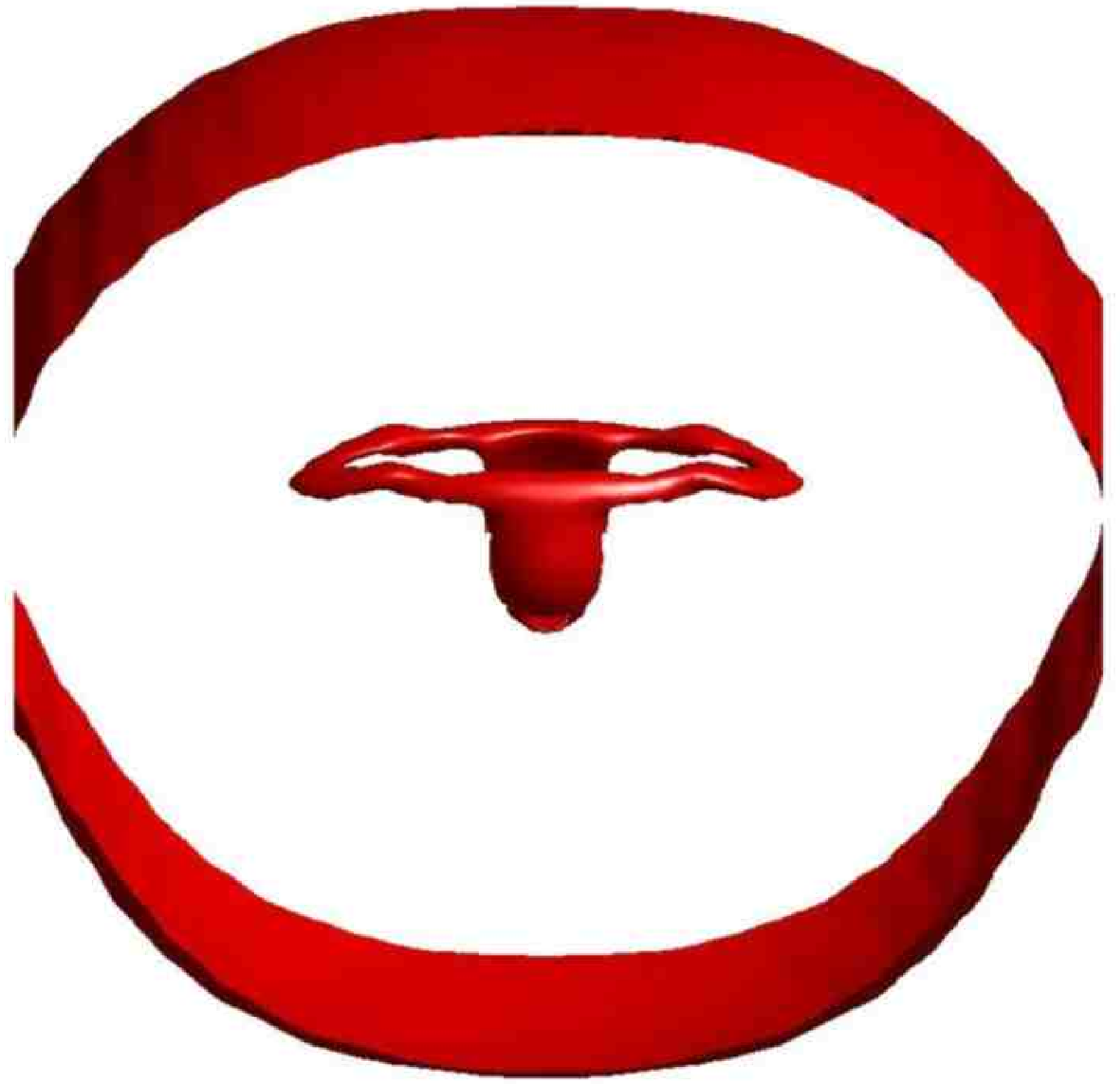}
\includegraphics[width= 0.49\columnwidth]{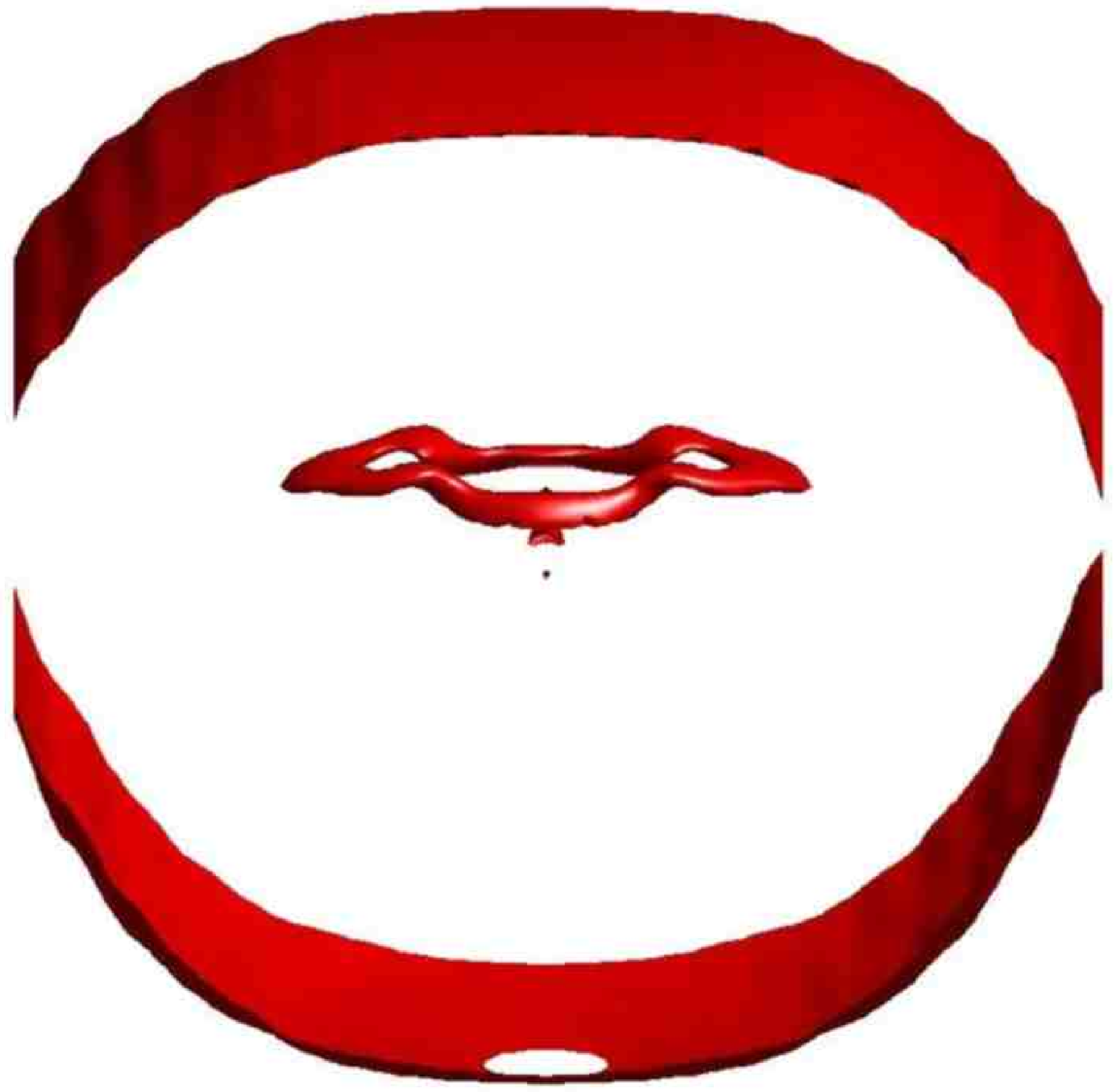}
\includegraphics[width=0.49\columnwidth]{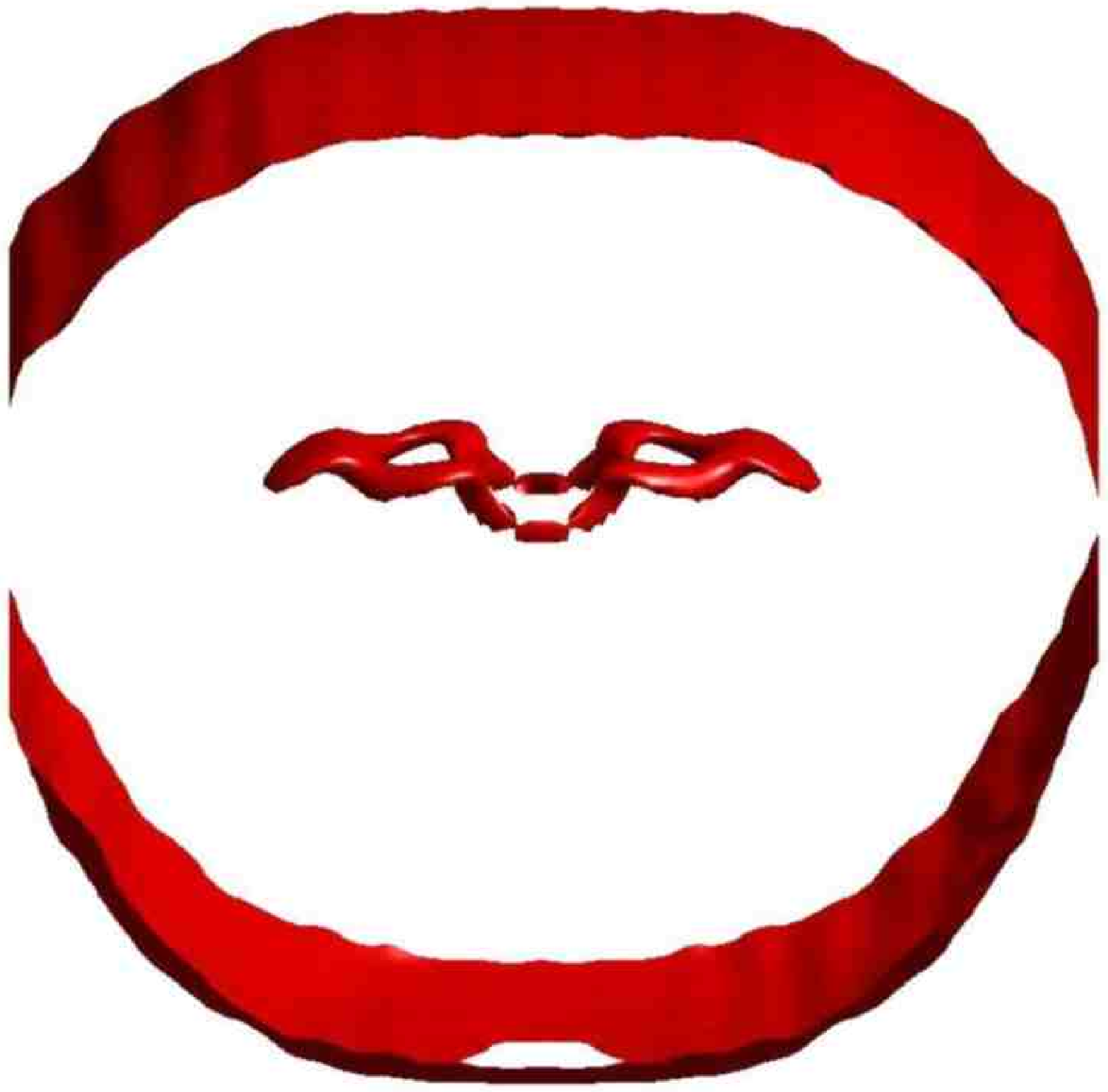}
\includegraphics[width=0.49\columnwidth]{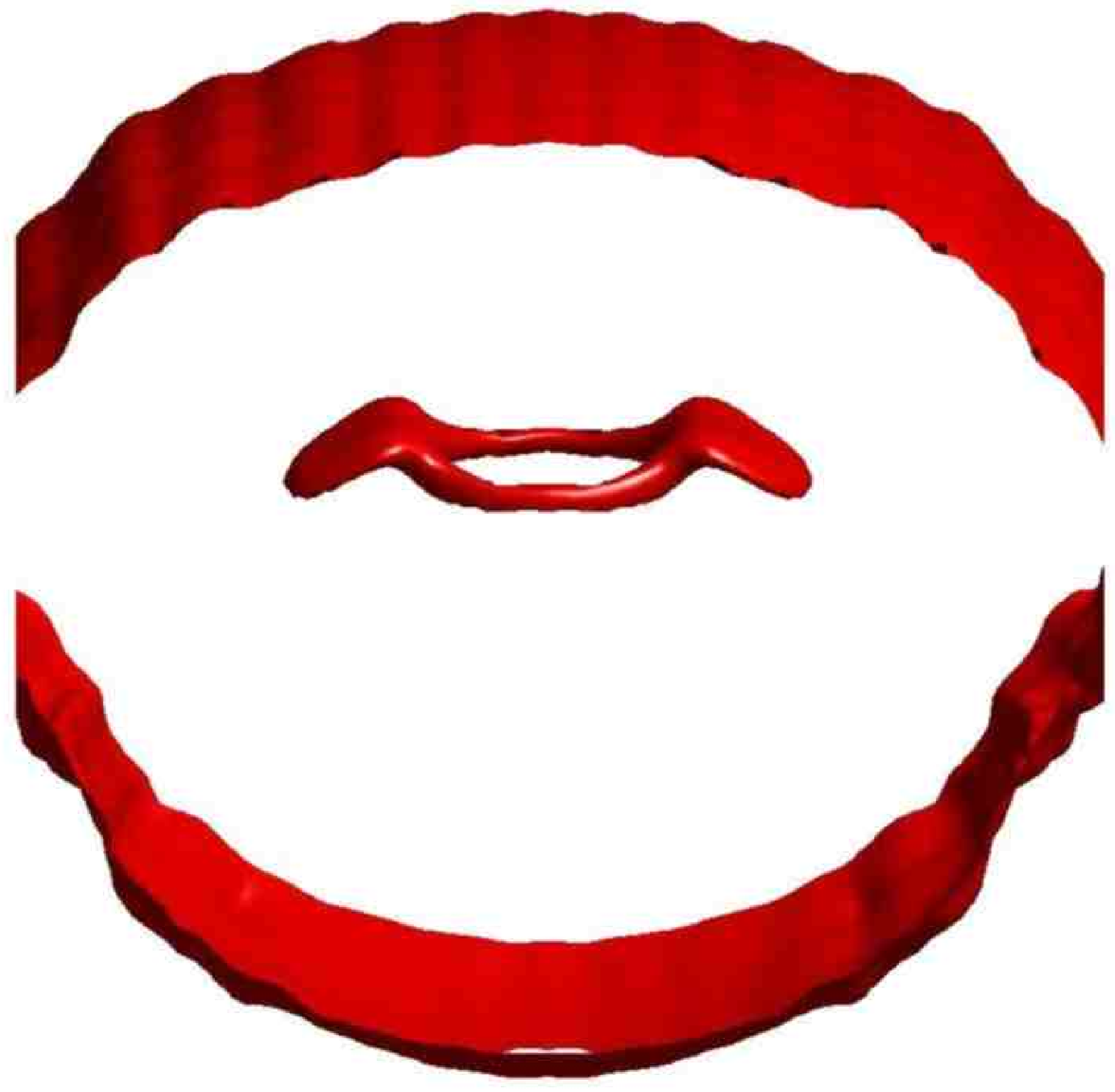}
\vspace{-0.3cm} \caption{The same system as in Fig.~\ref{ring1}, but
the initial state of the real time evolution was obtained by means
of propagating the wave functions in imaginary time for
$t\omega=0.02$. The resulting reconnection process is more violent
than the one displayed in Fig.~\ref{ring1}. We show the isosurface
plots at times $t\omega=0.1,0.2,0.3,0.4,0.5,0.6,0.8,1.0$.}
\label{ring2}
\end{figure}

In Figs.~\ref{ring1} and~\ref{ring2} we show the snapshot images of
the real time evolution of the two vortex rings after they were
prepared by means of the Rabi coupling, as explained in the previous
section. The images were obtained by numerically solving the coupled
GPE (\ref{gpe}) for the trapped two-component BEC in the absence of
the coupling fields $\Omega$. The integration was performed using
the split-step method \cite{JAV04} on a spatial grid of $128^3$. The
3D mean-field dynamics corresponds to the parameters of $^{87}$Rb
experiments, as described in the previous section. The number of
atoms in the two components $N_2\simeq 0.86N$ and $N_1\simeq 0.14N$,
where the total atom number $N=430l/(4\pi a_{12})$.

\begin{figure}
\includegraphics[width=0.8\columnwidth]{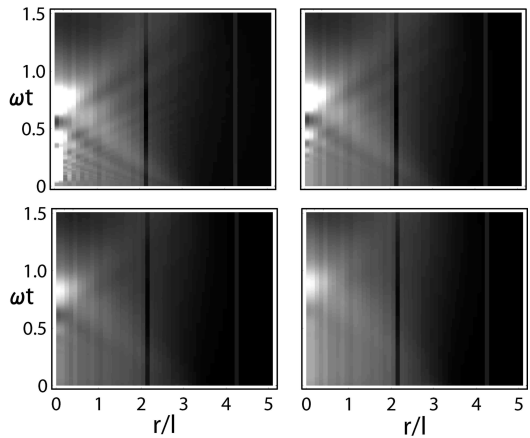}
\vspace{-0.3cm} \caption{The BEC energy density $\epsilon_t(r,t)$
averaged over the spherical angles during the reconnection process.
The initial imaginary time evolution $\omega t=0$ (top left),
$\omega t=0.02$ (top right), $\omega t=0.06$ (bottom left), and
$\omega t=0.1$ (bottom right). The white color indicates a high
value of the energy density. The sound waves first propagate towards
the trap center when the vortex cusp is generated as the two rings
merge. The energy is then radiated outwards and is transformed to
surface excitations. The straight vertical stripes are an artifact
of the numerical averaging procedure over the spherical angles.}
\label{ene1}
\end{figure}

\begin{figure}
\includegraphics[width=0.8\columnwidth]{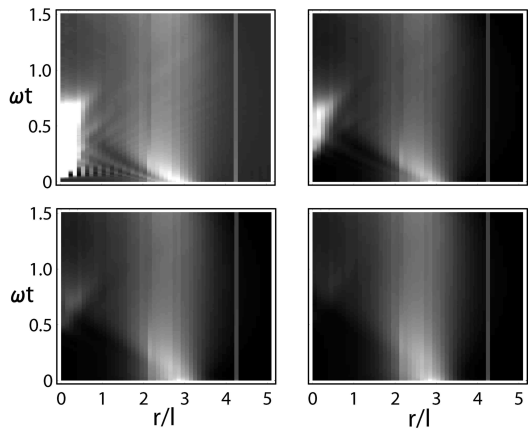}
\vspace{-0.3cm} \caption{The energy density $\epsilon_1(r,t)$
averaged over the spherical angles, depending only on the density of
the BEC component containing the vortex ring. We show the same cases
as in Fig.~\ref{ene1}. In the second and the third figure one may
clearly recognize the energy of the emitted sound pulses in the
fifth and the sixth images of Figs.~\ref{ring1} and~\ref{ring2}.}
\label{ene2}
\end{figure}

In Fig.~\ref{ring1} the initial state, obtained from the numerical
integration of the Rabi coupling, was relaxed by means of imaginary
time evolution of duration $t\omega=0.06$ before the displayed real
time evolution. In Fig.~\ref{ring2} the initial imaginary time
evolution was $t\omega=0.02$ and the merging process of the two
rings is more violent. During the imaginary time evolution we
separately normalized the both wave functions $\psi_1$ and $\psi_2$,
corresponding to the separate conservation of the atom number in
each BEC component in the absence of the coupling field. We also
show the total condensate energy density $\epsilon_t$, averaged over
the spherical angles,
\beq
\epsilon_t(r)\equiv {1\over4\pi} \int d\varphi\,
d(\cos\theta)\sum_{ij=1,2}\psi_i^*\big(H_0+{\kappa_{ij}\over2}|\psi_j|^2\big)\psi_i\,,
\eeq
as a function of the BEC radius and evolution time in
Fig.~\ref{ene1} for the cases shown in Figs.~\ref{ring1}
and~\ref{ring2}, as well as for cases without the initial imaginary
time evolution and for the imaginary time propagation of duration
$t\omega=0.1$. In Fig.~\ref{ene2} we show the contribution
$\epsilon_1$ to the energy density which only depends on the atom
density in the component $|1\>$ containing the vortex rings:
\beq
\epsilon_1(r)\equiv {1\over4\pi} \int d\varphi\,
d(\cos\theta)\,\psi_1^*\big(H_0+{\kappa_{11}\over2}|\psi_1|^2\big)\psi_1\,.
\eeq

In Figs.~\ref{ring1} and~\ref{ring2} the vortex rings are attracted
to each other as their motion is also bent downwards. It should be
emphasized that in the two-component BEC the vortex rings are
initially static, there is no thermal atom component, and, after the
preparation of the initial state, the dynamics is purely Hamiltonian
with conserved total energy and no added dissipation term. In order
to vortex rings to merge, they need to lose energy. In
Figs.~\ref{ene1}-\ref{ene2} we can clearly identify the released
excess energy as sound, radiated outwards and transformed into
surface excitations. It is the turbulent dynamics triggered by the
reconnection process which locally produce an effective dissipation
mechanism for the vortex configuration to relax. We can enhance the
sound emission and the reconnection dynamics by considering a more
rapid imprinting process. It is interesting to compare the merging
vortex rings to the formation of a vortex lattice in a rotating BEC.
In the latter case turbulent dynamics is triggered by the rotating
potential and the vortex lattice can locally relax in the classical
mean-field dynamics, even at $T=0$ and without any explicit damping
term \cite{LOB04}.

The reconnection process in Fig.~\ref{ene1} begins with a
significant concentration of energy towards the trap center, as
energy is propagated towards $r=0$. The vortex motion and the vortex
reconnection process excite Kelvin modes, resulting in a gradually
decreasing vortex line length. The Kelvin mode excitations of the
vortex cores are clearly observable in the first snapshot images of
Figs.~\ref{ring1}-\ref{ring2}. The generated vortex cusp at the
reconnection region detaches from the merging vortex rings and is
emitted as sound radiation. In the process the vortex energy is
converted into a sound pulse which propagates away from the vortex.
We observe large emission of vortex rings and/or rarefaction waves
\cite{BER05} at the reconnection region, e.g., in Fig.~\ref{ring1}
at time $t\omega=0.6$ and in Fig.~\ref{ring1} around $0.5\leq
t\omega\leq 0.6$. The emission events can also be seen in
Fig.~\ref{ene2} as high energy regions close to the origin. The
emitted sound energy results in strong surface excitations of the
BEC (displayed in the last two images of Figs.~\ref{ring1}
and~\ref{ring2}) with the fastest sound pulses (shortest wavelength
phonons) reaching the surface first.

\section{Using ultra-slow light in the studies of vortex
collisions} \label{slow}

\subsection{Dissipation in vortex pair collisions}

One of the fascinating aspects of vortices in current BECs is the
fact that the relatively small nonlinearity leads to many finite
compressible effects.  In geometries where the system is essentially
two dimensional, due to tight confinement in one direction (we will
label $y$ here), vortices act as point particles.  In an
incompressible fluid, they are indestructible particles which
interact via conservative potentials, moving always according to the
velocity fields created by neighboring vortices.  For example, two
vortices of like charge will orbit each other at some constant
radius, while oppositely charged vortices will form `pairs' which
travel along straight lines maintaining the same distance between
them $d$ \cite{FET99}.  In either case, the velocities are inversely
proportional to the distance to the neighboring vortices. However,
when the distances between vortex pairs become comparable to the
healing length (which is the characteristic size of vortex cores),
this picture breaks down. For example, in the case of co-rotating
vortices of like charge, the pair will emit sound radiation and
slowly drift to larger distances, lowering their energy
\cite{LUN00}.  The radiation occurs due to acceleration of the
vortices as they undergo circular motion. Finite compressibility
effects have also been important in explaining discrepancies between
experimental and theoretical descriptions in excitations of vortex
lattices \cite{Baym03}.

The dynamics of multiple vortex configurations, particularly ones
containing vortices of both circulations, can be particularly rich.
For example, scattering between vortex pairs have been studied
\cite{KOV02} as well as the existence and behavior of soliton-like
structures in 2D geometries, and how they can dynamically form from
closely spaced vortex pairs \cite{HUA03}. However, there has been
little work on cases where the dissipation due to compressibility is
significant and vortex annihilations occur. The dissipation is
analogous to the length shortening and sound emission which occurs
with colliding rings in 3D geometries studied above and in
Refs.~\cite{KOP96,adams}.

Here we show that the recently developed technique of the light
roadblock, based on ultra-slow light propagation in atomic BECs, may
provide a method of preparing point vortices and systematically
investigating their collisions in strongly anisotropic
`pancake-shaped' atom clouds. We review the ultra-slow light
propagation in BECs and the creation technique by which systems
containing multiple vortices of each charge can be engineered,
allowing study of these interesting phenomena which have remained
unexplored experimentally to now. In particular, we show how it is
possible to induce vortex pair scattering, in a regime where the
dissipation due to finite compressibility is quite apparent.  We
then study such scattering in more detail to obtain new quantitative
results on the role of dissipation in such collisions.

\subsection{Ultra-slow light propagation and the light roadblock}

Slow light is based on electromagnetically-induced transparency
(EIT) \cite{EIT}, a quantum interference effect that permits the
propagation of light through an otherwise opaque medium. A coupling
laser, resonant with some initially unoccupied hyperfine state
$|2\>$ and an optically excited state $|3\>$, is used to create the
interference resulting in transparency (vanishing resonant
absorption), a very large dispersion, and ultraslow group velocity
of a probe pulse, resonant with an initially occupied state $|1\>$
and $|3\>$. Probe light pulses have been slowed down and spatially
compressed (compared to their free space propagation values) by up
to eight orders of magnitude \cite{Nature1,KAS99,BUD99}.  As the
probe propagates through the BEC, it puts the atoms into spatially
dependent superpositions of $|1\>$ and $|2\>$ with the relative
phase and amplitude of the wavefunctions (the atomic coherence)
reflecting the amplitude and phase pattern of the input probe.

Combining this technique in BECs with the recently developed
experimental technique of the `light roadblock' \cite{DUT01},
whereby a fast spatial variation of the coupling field is
introduced, provides an efficient method of inducing phase
singularities on matter waves. Sharp density defects are prepared by
abruptly perturbing the light propagation. In anisotropic 2D trap
configurations the created defects evolve in a predictable way to
solitonlike phase kinks and subsequently then to systems of multiple
point vortices (of both circulations) whose collisions may be
investigated in detail.  Here we focus on the case of a strongly
anisoptropic trap, leading to essentially 2D dynamics. In
Ref.~\cite{DUT01}, the geometry was truly 3D and in
Ref.~\cite{GIN05} the 3D aspects of the resulting topological
structures in a similar system were investigated.

We specifically consider the BEC occupying two hyperfine levels
$|1\>$ and $|2\>$. These are connected, respectively, to a common
excited state $|3\>$ by orthogonally propagating, resonant laser
fields: the $+z$ propagating probe $\Omega_{13}$ and $+y$
propagating coupling $\Omega_{23}$, each of wavelength $\lambda$.
The excited state $|3\>$ decays at $\Gamma$, forming a $\Lambda$
three-level structure. In the slowly varying envelope (SVE)
approximation for the light fields $\Omega_{i3}$ (with the rapid
phase rotation at the optical frequencies and optical wavenumbers
factored out) the Maxwell's equations read \cite{processing}:
\begin{align}
\label{eq:formalismPC} \left( \frac{\partial}{\partial z}+
\frac{1}{c}\frac{\partial}{\partial t} \right) \Omega_{13}  =& -
f_{13} \sigma N[\Omega_{13} |\psi_1|^2\nonumber
\\ &+\Omega_{23}\psi^*_1\psi_2
e^{-i k (z-y)}],
\end{align}
\begin{align}
\label{eq:formalismPC2}\left( \frac{\partial}{\partial y}+
\frac{1}{c}\frac{\partial}{\partial t} \right) \Omega_{23}  = &-
f_{23} \sigma N[\Omega_{23} |\psi_2|^2\nonumber
\\&+\Omega_{13}\psi^*_2\psi_1 e^{i k (z-y)}].
\end{align}
\noindent Here $N$ is the initial total number of BEC atoms, $k=2
\pi/\lambda$, $f_{i3}$ are dimensionless oscillator strengths, and
$\sigma\equiv 3 \lambda^2/2 \pi$ is the resonant cross-section. In
Eqs.~(\ref{eq:formalismPC}-\ref{eq:formalismPC2}) we have
adiabatically eliminated $\psi_3$ by assuming that the dynamics of
the internal atomic degrees of freedom and the light are much faster
than the external dynamics \cite{JAV95b,processing}.  The BEC
wavefunctions $\psi_1,\psi_2$ evolve according to generalized GPEs:
\begin{align}
\label{eq:formalism12} i \hbar  \dot\psi_{i} = &\big(H_0+ \sum_k
\kappa_{ik} | \psi_k |^2\big)
\psi_{i}\nonumber\\&-{i\hbar\over\Gamma}\big[
|\Omega_i|^2\psi_i+\Omega^*_i\Omega_j\psi_j e^{-i k
(\epsilon_i-\epsilon_j)}\big] \,,
\end{align}
where $i=1,2, \, j \not= i$ and $\epsilon_1=z,\epsilon_2=y$. The
last term in Eq.~(\ref{eq:formalism12}) results in both coherent
exchange between $|1\rangle,|2\rangle$ as well as absorption into
$|3 \rangle$. The phase factors assure that atoms coherently coupled
into $|2 \rangle$ experience a two-photon momentum recoil relative
to the nearly stationary $|1 \rangle$ component.  In our model,
atoms which populate $|3 \rangle$ and then spontaneously emit are
assumed to be lost from the BECs.

If one were to input a weak ($|\Omega_{13}| \ll |\Omega_{23}|$)
free-space pulse of some length $L_z$ into a BEC, simultaneous
solution of Eqs.(\ref{eq:formalismPC}-\ref{eq:formalism12}) reveals
that the pulse is compressed to a length $(V_g/c)L_z$, where the
group velocity is $V_g=|\Omega_{23}|^2/\Gamma f_{13} \sigma N
|\psi_1|^2$, as it propagates through the BEC.  Any phase and
amplitude features along the longitudinal direction are similarly
compressed while the transverse features are unaffected. While the
usual description of slow light propagation depends on the weak
probe assumption ($|\Omega_{13}| \ll |\Omega_{23}|$), the basic
transfer of amplitude and phase patterns works even when this
assumption is not strictly satisfied.  This regime was studied in
detail in Ref.~\cite{processing}.

Spatial modulation of the input coupling field along $z$ is then
used to vary the speed and length of slow light pulses as they
propagates through the BEC. In Ref.~\cite{DUT01} this variation was
accomplished by using a razor blade to block the coupling field from
the BEC in the $z>0$ region so near $z=0$ the coupling intensity
quickly varied from some large value to zero.  In this region, the
probe beam is slowed to zero group velocity and the length becomes
arbitrarily small. At the same time the probe to coupling intensity
ratio becomes large in this region, and so the fraction of atoms
coupled into $|2 \>$ becomes large. The result is a very narrow,
large amplitude, density defect in the original condensate internal
state $|1\>$, with the atoms in this region coupled into the other
stable internal state $|2\>$. Because of the large photon recoil in
the orthogonal geometry, atoms transferred to $|2\>$ leave the
defect region in $\alt 1$ ms, leaving an empty hole in a single
component BEC. This method is particularly useful for inducing
density features in BECs smaller than the experimental optical
resolution of the system, as the spatial compression at the
roadblock occurs due to the slow light propagation effects and not
focusing.  This is particularly valuable in creating topological
defects, since their characteristic length scale, the healing
length, is typically $< 1~\mu$m.  Numerical simulation of the
$^{23}$Na experiment in Ref.~\cite{DUT01} indicates that density
defects $\sim 2~\mu$m long (and nearly 100\% depth) were being
created even though the spatial variation of the coupling field
(which {\it is} limited by optical resolution) was about $\sim
10~\mu$m.  This interpretation was further supported by the fact
that the density defects of the features were unobservable in
in-trap absorption images but were clearly visible after 1~ms of
free expansion.

\subsection{Inducing vortex pair collisions with a `light
roadblock'}

\begin{figure}
\includegraphics[width=0.98\columnwidth]{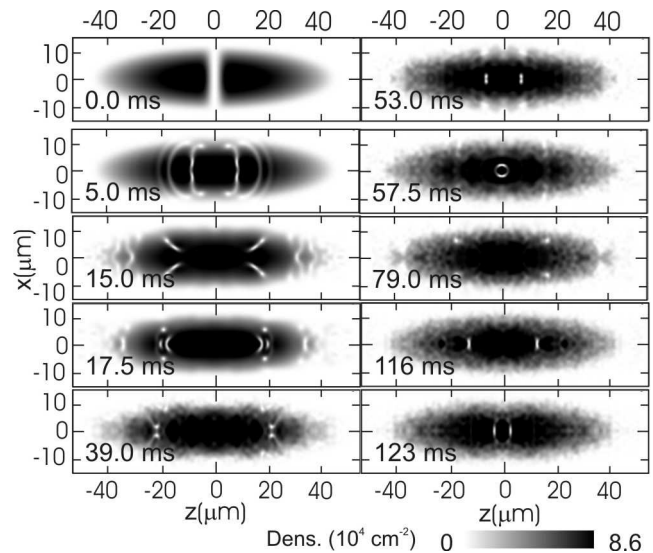}
\caption{The evolution of a $3~\mu$m ($1/e$ half-width), full depth
defect in a sodium BEC at the times indicated.  Qualitative features
of the evolution are explained in the text.  The density scale
refers to the normalized 2D density $|\tilde\psi(x,z)|^2$.}
\label{defectEvol}
\end{figure}

In the original experiment \cite{DUT01} the imprinted density
defects were seen to subsequently break up into a series of a
solitons, due to the superfluid analog of a shock waves.  These
solitons, in turn, broke up into pairs of vortices via the snake
instability \cite{AND01,KP70,FED00}. We show an example of how this
comes about in Fig.~\ref{defectEvol}.

In this example we consider a BEC with repulsive interactions in a
trap with much stronger confinement along the $y$ axis, with the
trap frequencies satisfying $\omega_y\gg\omega_{x},\omega_z$, so
that the 2D limit of the GPE is valid. We assume that the defect has
been created using the light roadblock technique, as explained
earlier. Moreover, after the light fields have been turned off, we
assume that the BEC component $|2\>$, which is initially filling the
density defect in $|1\>$, is instantaneously coupled out of the trap
due to the photon-induced recoil, as in the original experiments
\cite{DUT01}. Then the BEC dynamics approximately  follows from the
single-component 2D GPE for $\tilde\psi(x,z)$, when we write
$\psi_1(\rv)\simeq \phi(y) \tilde\psi(x,z)$, in terms of the ground
state harmonic oscillator wave function $\phi(y)$ along the $y$
axis:
\beq
i \hbar {\partial \tilde\psi\over \partial t} = \big[
-{\frac{\hbar^2}{2m}}{\bf \nabla}^2+{\frac{m}{2}} (\omega_z^2
z^2+\omega_x^2 x^2) + \kappa_{2D} |\tilde\psi|^2 \big] \tilde\psi\,,
\label{gpe2d}
\eeq
where $\kappa_{2D}=\kappa_{11}/l_y \sqrt{2\pi}$,
with $l_y=\sqrt{\hbar/m\omega_y}$. The additional density-dependent
contribution to the scattering length in 2D is negligible when
$\sqrt{2\pi}l_y/a\gg \ln{(8\pi^{3/2}l_y n_{2D} a)}$ \cite{PET00},
where $n_{2D}=N|\tilde\psi|^2$ denotes the 2D density. In the
numerics, we choose $\omega_x/\omega_z\simeq3.81$ and the
nonlinearity $\kappa_{2D} = 2360 \, \hbar \omega_z l_z^2$
($l_z=\sqrt{\hbar/m\omega_z}$). The numerical results depend on the
dimensionless nonlinearity
$\bar\kappa_{2D}\equiv\kappa_{2D}/\hbar\omega_z
l_z^2=2\sqrt{2\pi}Na/l_y$, so our results are unchanged in any
rescaling of the parameters which maintains constant $Na/l_y$ and
$l_z/l_x$, provided that the $y$ confinement remains large enough
for the 2D GPE (\ref{gpe2d}) to be valid.

We then remove atoms from the relaxed ground state to create a
narrow ($\sim 0.7l_z$ half width) density hole, with 100\% of the
density removed along the central $z=0$ axis. The GPE (\ref{gpe2d})
was propagated with a Crank-Nicholson grid with an equal spaced grid
(300 points in $x$ and 600 in $z$), with time steps $1.65\times
10^{-7}/\omega_z$. Both were varied to assure the results were not
effected.

To make the connection with experimental parameters, we quote our
results in terms of $^{23}$Na parameters, which has a scattering
length $a\simeq2.75$~nm, and choose a trap $\omega_z=2
\pi\times21$~Hz, giving $l_z\simeq4.6~\mu$m and defect size
$\sim3~\mu$m. The ground state BEC in this trap has a central peak
density of $n_{3D}=1.5 \times 10^{-14}~\mathrm{cm}^{-3}$ and the
superfluid healing length at the cloud center is $\zeta\sim \hbar/(2
m \kappa_{2D} |\tilde\psi|^2)^{1/2}\simeq 0.4~\mu$m.

In the limit of small-amplitude and large wavelength, the imprinted
defect would split into two counter-propagating sound waves
\cite{AND97}. However, due to the nonlinearity of the excitation
(and the small length scale) \cite{DUT01} it here sheds off a series
of a gray solitons \cite{MOR97}, as seen in the snapshot at 5~ms. In
addition, the slowest, and deepest soliton has begun to curl due to
the snake instability, and vortices have begun to form. Because
angular momentum is not being imparted to the system, vortices are
formed in pairs of opposite charge. Investigation of the phase shows
all the vortices formed are singly quantized.

At 15~ms each quadrant contains a line of three vortices
(alternating in sign along each line).  Note that they are closely
spaced here so their density profiles overlap, and the
identification of the vortices is done by finding singularities in
the phase profile. Each vortex induces a circular velocity field
about it, with the magnitude falling off as the inverse of the
distance from the vortex center, and the direction determined by its
charge. The motion of each vortex is governed by the velocity field
of the condensate, which is typically dominated by the velocity
field induced by nearby vortices. In our example, this causes each
of these lines to spin in pin-wheel fashion. In the 17.5~ms frame,
this motion has induced collisions between pairs of vortices (from
$x>0$ and $x<0$), resulting in annihilations of the vortices, with
the energy carried off in sound waves, (seen as curved density
defects) which quickly propagate towards the condensate edge and
break up. What remains are four vortices, as seen at 39~ms.  At
53~ms these vortices form two vortex pairs, that, due to the small
distance between each pair, move very quickly towards the condensate
center, on a collision course.  Upon close approach with the other
pair, the vortices make very sudden 90 degree terms and switch
partners. Associated with this collision is a rather noticeable
sound pulse resulting in a ring like shape, seen at 57.5~ms.
However, the four vortices survive and due to their interaction with
the BEC edge, are sent around the edge of the BEC (79~ms) and
eventually back along the same collision course (116~ms).  This time
the pairs are so close together that their density profiles greatly
overlap (compared with the first approach, 53~ms). Upon the second
collision, at 122~ms, rather than make the 90 degree turn, the
vortices are destroyed (no phase singularities are visible after the
collision) and the density defects propagate through each other and
continue along the $z$ axis (123 ms).

This sequence of vortex dynamics demonstrates a myriad of
interesting features associated with the finite compressibility of
the superfluid.  In the limit of no compressibility, there is no
sound emission and the vortex dynamics is completely conservative.
In the absence of effects associated with the background condensate
density, a vortex pair collision would look much like the collision
at 57.5~ms, but with no associated sound emission, and with the size
of the outgoing vortex pairs $d$ (at 57.5~ms), the same as the
incoming size (53.0~ms). However, due to the finite speed of sound,
the vortices can exchange energy with phonons, dissipating some of
their energy, as in the first collision of pairs at 57~ms, and in
some cases annihilating, as in the second collision at 122~ms. In
fact, the dissipation experienced during the first collision is what
ultimately lowered the energy of each vortex pair, causing the pairs
to to be even smaller as they entered the second collision.

We note that the number of vortices created with the light roadblock
and their subsequent dynamics can be be controlled both by the size
of the defect (which is ultimately controlled by the number of
photons in the input probe pulse) and anisotropy in the $x-z$ plane.
For example, in Ref.~\cite{DUTEPN} it was shown how a more circular
shaped BEC leads to the production of many more vortices in each
soliton.  This geometry would be more favorable for minimizing
effects due to BEC density variations.   A sufficiently elongated
trap would produce only two per soliton, and may be more useful for
tightly controlling the vortex trajectories and thus ensuring a
direct collision is induced.

\subsection{Numerical study of dissipation in vortex pair collisions}

To study this collision induced dissipation more systematically, we
performed a series of numerical calculations of vortex pair
collisions in a homogenous BEC.  This was accomplished by first
relaxing a condensate in a trap consisting of a large square flat
potential surrounded by steep trapping walls.  The product of the
homogenous normalized 2D density and the nonlinearity
$\kappa_{2D}|\tilde\psi^{(0)}|^2$ determine the characteristic
length scale (the healing length) $\zeta\simeq\hbar/(2 m
\kappa_{2D}|\tilde\psi^{(0)}|^2)^{1/2}$, speed of sound
$c_0=(\kappa_{2D}|\tilde\psi^{(0)}|^2/m)^{1/2}$, and time scale
$t_0=\hbar/\kappa_{2D}|\tilde\psi^{(0)}|^2$.  The 2D GPE
(\ref{gpe2d}) can then be renormalized into dimensionless form and
we quote our results for this section in these units.   Our
homogenous trap region was chosen to by $78 \zeta \times 78 \zeta$
and for numerical calculation was broken up into a grid with spacing
$0.21 \zeta$.

After relaxation we then impose four vortices: positive charge ones
(clockwise velocity circulation) at $(x=7.4 \zeta, z=14.7 \zeta)$
and $(-7.4 \zeta,-14.7 \zeta)$ and negative ones at $(-7.4
\zeta,14.7 \zeta)$ and $(7.4 \zeta,-14.7 \zeta)$.  In each case we
impose the $e^{\pm i \phi}$ phase and a numerical estimate for the
density profile \cite{FET99}.  This configuration is then relaxed
further.  As it relaxes, each vortex pair shrinks ({\it i.e.} the
vortices drift towards $x=0$ and towards each other), as bringing
vortices of opposite charge together lowers the overall energy. We
then stop this relaxation at various times and begin a real time
evolution of the GPE. The circulations of the vortices are such that
the pairs move towards each other and collide.  In this way we can
systematically study how vortex pairs of various initial sizes
$d_\mathrm{in}$ act upon collision.  For both relaxation and
propagation of the GPE we used a Crank-Nicholson algorithm, with
time steps $0.01 t_0$.

\begin{figure}
\includegraphics[width=0.75\columnwidth]{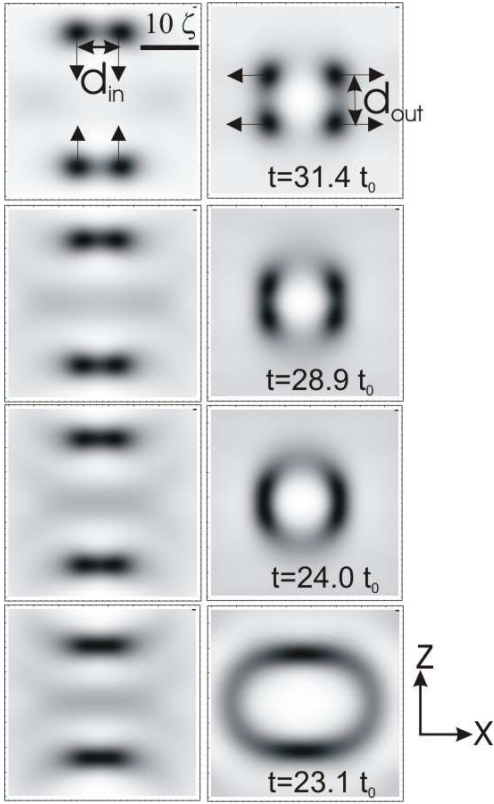}
\caption{Examples of vortex pair collisions, with each row showing
two snapshots of a separate simulation with different initial pair
distances $d_\mathrm{in}$.  Note that here white indicates high
density (in contrast to Fig.~\ref{defectEvol}). The left (right)
hand column shows the vortices before (after) the collision of the
pair.  The initial distances of the pairs are, from top to bottom,
$d_\mathrm{in}/\zeta=7.0, \, 5.4, \, 4.6,$ and $3.3$. The arrows in
the top row indicate the vortex direction. } \label{collEx}
\end{figure}

Several examples of collisions are shown in Fig.~\ref{collEx}. In
the top example, $d_\mathrm{in}=7.0 \zeta$ and the vortices travel
along the paths indicated by the arrows. Upon collision the vortices
turn 90 degrees and switch partners.  There is a barely perceptible
sound wave upon the collision. This sound radiation is due to the
sudden acceleration experienced by the collision.  In the second row
$d_\mathrm{in}=5.4 \zeta$, and a noticeably larger sound wave is
generated (due to the larger acceleration experienced by a smaller,
and therefore faster, pair). Even though the density profiles of the
pairs significantly overlap, the phase singularities remain in the
wavefunction $\psi_1$ after the collision as the vortices travel
out.  This example is analogous to the first collision in
Fig.~\ref{defectEvol} (57~ms).  In the third row, $d_\mathrm{in}=4.6
\zeta$. In this case, the sound wave is bigger still, and the
outgoing vortices have completely merged. Examining the phase
indicates also that the phase singularities annihilate shortly after
the collision. These structures bear resemblance to the `lumps'
studied in \cite{HUA03}.  In the bottom case, $d_\mathrm{in}=3.3
\zeta$, the vortices annihilate early on in the collision, before
the trajectories of the vortices have been significantly altered,
and the remaining density waves simply pass through each other. This
is similar to what happened in the second collision of
Fig.~\ref{defectEvol} (122~ms). It is interesting to observe how the
sound wave energy emission pattern is strikingly different between
the third and fourth cases of Fig.~\ref{collEx}, with the the energy
emission primarily horizontal in the former and vertical in the
latter.  We also observed an intermediate case ($d_\mathrm{in}=4.2
\zeta$), in which the sound energy propagated away with near perfect
cylindrical symmetry.

\begin{figure}
\includegraphics[width=0.98\columnwidth]{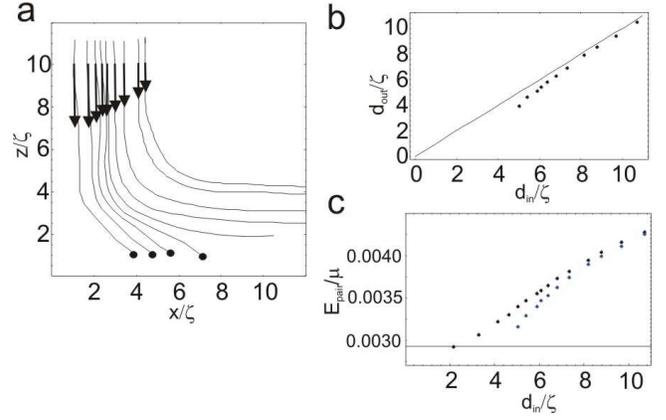}
\caption{\textbf{(a)} The trajectories (specifically of the phase
singularities) followed by the upper-right vortex for several
different input sizes $d_\mathrm{in}$. The arrow lengths are
proportional to the observed velocity of the pairs at $z=10 \zeta$
(with the longest arrow corresponding to $0.48 \, c_0$). In four of
the cases the vortices annihilated at the locations indicated with
the dots.  In the other cases, the phase singularities survived the
collision. \textbf{(b)} For a series of cases, we plot the incoming
pair size $d_\mathrm{in}$ (calculated when they are at $z=\pm 10
\zeta$) versus the outgoing size $d_\mathrm{in}$ (when they reach
$x=\pm 10 \zeta$). \textbf{(c)} The energy of the vortex pairs
versus the input distance $d$ (black dots), calculated by taking
half the energy of the two vortex pair configuration above the
ground state energy. The horizontal line indicates the point at
which the phase singularities disappear. The blue dots then indicate
the pair energies, calculated via the observed $d_\mathrm{out}$.}
\label{InOut}
\end{figure}

In Fig.~\ref{InOut}(a) we plot a series of trajectories followed by
the vortices during these pair collisions (due to the symmetry, we
only plot the vortex in the $x>0,z>0$ quadrant in each case).
Initially all are following a trajectory almost purely in the $-z$
direction, with velocities proportional to the arrow lengths in the
figure. The input size of the vortex pairs $d_\mathrm{in}$ is twice
the initial $x$ position. For sufficiently large vortex pairs $d \gg
\zeta$ the velocity varies as $1/d$, in agreement with the theory of
an incompressible fluid, however, as $d$ becomes comparable to
$\zeta$ the velocity begins to saturate, reaching a value $0.48 \,
c_0$ for the smallest pair we observed, $d_\mathrm{in}=2.18 \zeta$.
The phase singularities disappeared when the GP relaxed the vortex
pair sizes to smaller than this. As each vortex approaches the
counter-propagating pair near the origin, it begins to feel the
pair's influence and turn to the $+x$ direction, eventually forming
a new pair with its counterpart in the other pair. Again, for
$d_\mathrm{in} \gg \zeta$ vortices, the system is conservative and
the new vortex pairs propagate out with the same size
$d_\mathrm{out}=d_\mathrm{in}$. However, as we saw in
Fig.~\ref{collEx}, large sound waves emit energy for smaller vortex
pairs, and as a result, the outgoing pairs have lower energy and
thus $d_\mathrm{out}<d_\mathrm{in}$. The outgoing versus incoming
sizes are plotted for a series of cases in Fig.~\ref{InOut}(b) and
the trend towards more dissipation for smaller $d_\mathrm{in}$ is
clearly seen.  For the four smallest $d_\mathrm{in}$ plotted in
Fig.~\ref{InOut}(a), the dissipation is sufficient to induce an
annihilation event at the locations indicated by the dots (thus
these cases are not plotted in Fig.~\ref{InOut}(b)).

Since the fundamental mechanism here is energy dissipation via sound
emission, it is useful to analyze the numerical simulations from an
energy point of view.  To do this, we calculate the energy of our
relaxed wave function at the beginning of each simulation. By
subtracting from this the ground state energy (without vortices) and
halving the result (since there are two vortex pairs), we obtain a
vortex pair energy as a function of size $E_\mathrm{pair}(d)$. This
quantity is plotted for each $d_\mathrm{in}$ as black dots in
Fig.~\ref{InOut}(c). The horizontal line indicates the point at
which no phase singularity was observed ($d<2.18 \zeta$). We then
plot, using the observed outgoing size $d_\mathrm{out}$ and our
numerically calculated energy dependence $E_\mathrm{pair}(d)$, the
outgoing pair energy and plot it for versus $d_\mathrm{in}$ as blue
dots. The difference between the two black and blue dots thus
indicates the energy carried off by phonons in each case.  For the
four smallest incoming pairs, the energy dissipation was large
enough to that the pair energy fell below the horizontal line and
the vortex phase singularity was eliminated.

\section{Conclusions}

We investigated methods for imprinting desired patterns of phase
singularities on atomic matter waves by means of em fields inducing
transitions between different hyperfine levels of the atom as well
as by means of controlling the ultra-slow light propagation inside
the atomic cloud. By appropriate field superpositions one may
construct phase singularities around the node points of the em field
amplitude which can then be imprinted on the BECs using coherently
driven atomic transitions. In particular, we showed that a simple
superposition of two Gaussian laser beams, which exhibit a different
beam waist, is sufficient to imprint a phase singularity forming a
closed circular loop with a $2\pi$ phase winding around the loop.
This provides a controlled method of creating vortex rings on atomic
superfluids, where the size, the position, and the orientation of
the ring can be engineered by modifying the parameters of the beam
superposition. Moreover, we showed that the imprinting techniques
can be generalized for imprinting vortex ring pairs and multi-vortex
systems which are especially interesting in the studies of vortex
reconnections. The experimental technology of diffractive optical
components, such as optical holograms, may be particularly useful in
generating the light beams with the desired phase and the amplitude
profiles. We investigated in detail one specific example of vortex
reconnections resulting from the imprinted phase profile by
considering two merging vortex rings in a trapped, two-component
atomic BEC. We used the parameters of $^{87}$Rb experiments and
calculated the energy distribution in the atom cloud during the
reconnection process.

In addition to direct imprinting of phase singularities on the BECs,
we also explored the recently developed experimental technique of
preparing defects by means of abruptly distorting the ultra-slow
light propagation in atomic BECs with `light roadblocks'. We showed
that in a tightly-confined 2D configuration this technique is
particularly useful in the studies of point vortex collisions.
Similarly to the reconnection dynamics of the vortex rings in an
isotropic trap, we explored the role of sound emission in the
collision processes of the point vortices and showed how the binding
energy of the pair can be dissipated during close range collisions.
Sufficiently large energy dissipation leads to the interesting
phenomenon of vortex annihilation.

Our results point the way towards a number of experiments that can
be implemented with present day cold-atom and laser technology, and
that could allow novel and controlled experimental investigation of
complicated vortex interactions.

\acknowledgments {J.R. acknowledges discussions with James Anglin
and Mark Dennis and financial support from the EPSRC and NSF through
a grant for the Institute for Theoretical Atomic, Molecular and
Optical Physics at Harvard University and Smithsonian Astrophysical
Observatory. Z.D. acknowledges NIST Gaithersburg, where some of this
work was carried out, and discussions there with Jamie Williams.}

\end{document}